\documentclass[conference,hidelinks]{IEEEtran}
\usepackage{amsmath,amsfonts}
\usepackage{algorithmic}
\usepackage{fontawesome5}

\usepackage{array}
\usepackage[caption=false,font=normalsize,labelfont=sf,textfont=sf]{subfig}
\usepackage{textcomp}
\usepackage{stfloats}
\usepackage{hhline}
\usepackage{url}
\usepackage{verbatim}
\usepackage{graphicx}
\def\BibTeX{{\rm B\kern-.05em{\sc i\kern-.025em b}\kern-.08em
    T\kern-.1667em\lower.7ex\hbox{E}\kern-.125emX}}
\usepackage{balance}

\usepackage[table,dvipsnames]{xcolor}
\usepackage[shrink=25]{microtype}
\usepackage{tikz}
\usetikzlibrary[shadows]
\usepackage{ifthen}
\usetikzlibrary{calc}
\usetikzlibrary{arrows.meta}
\usetikzlibrary{external}
\usetikzlibrary{positioning,fit,calc}
\usepackage[mathscr]{eucal}
\usepackage{dsfont}
\usepackage{scalefnt}
\usepackage{amsmath,amssymb,amsfonts}
\usepackage{algorithmic}
\usepackage{graphicx}
\usepackage{textcomp}
\usepackage{listings}
\definecolor{backcolour}{rgb}{0.95,0.95,0.92}
\definecolor{arrowcolor}{RGB}{145,148,138}
\definecolor{weborange}{RGB}{255,165,0}
\definecolor{darkblue}{rgb}{0.0,0.0,0.6}
\definecolor{cyan}{rgb}{0.0,0.6,0.6}
\definecolor{deepred}{rgb}{0.6,0,0}
\definecolor{deepgreen}{rgb}{0,0.5,0}

\lstdefinelanguage{PAULA}{
  keywords={par, and, or, if, ifrt, in, out, program, variable, parameter},
  emph={matrix_vector_multiplication, mvm},
  emphstyle=\color{deepred},
  comment=[l]{//},
  commentstyle=\color{green!50!black},
  keywordstyle=\color{blue}
}

\lstdefinestyle{my_PAULA_style}{
  language=PAULA,
  backgroundcolor=\color{backcolour},
  breaklines=true,
  xleftmargin=1.5em,
  numbers=left,
  numbersep=0.5em
}

\lstdefinestyle{PAULA_style}{
  language=PAULA,
  breaklines=true
}

\lstdefinestyle{PseudoCode}{%
  keywords={for, do},
  breaklines=true,
  backgroundcolor=\color{backcolour},
  keywordstyle=\color{blue}
}

\lstdefinestyle{log}{%
  breaklines=true,
  backgroundcolor=\color{backcolour}
}

\lstdefinelanguage{json}{%
  basicstyle=\normalfont\ttfamily,
  breaklines=true,
  backgroundcolor=\color{backcolour},
  literate=
     *{0}{{{\color{deepred}0}}}{1}
      {1}{{{\color{deepred}1}}}{1}
      {2}{{{\color{deepred}2}}}{1}
      {3}{{{\color{deepred}3}}}{1}
      {4}{{{\color{deepred}4}}}{1}
      {5}{{{\color{deepred}5}}}{1}
      {6}{{{\color{deepred}6}}}{1}
      {7}{{{\color{deepred}7}}}{1}
      {8}{{{\color{deepred}8}}}{1}
      {9}{{{\color{deepred}9}}}{1}
      {:}{{{\color{deepred}{:}}}}{1}
      {,}{{{\color{deepred}{,}}}}{1}
      {\{}{{{\color{darkblue}{\{}}}}{1}
      {\}}{{{\color{darkblue}{\}}}}}{1}
      {[}{{{\color{darkblue}{[}}}}{1}
      {]}{{{\color{darkblue}{]}}}}{1},
}

\lstdefinestyle{myC}{%
  language=C,
  tabsize=2,
  keywordstyle=\color{blue},
  commentstyle=\color{green!50!black},
  stringstyle=\color{deepgreen},
  backgroundcolor=\color{backcolour},
  breaklines=true,
  breakatwhitespace=true,
  moredelim=[s][\color{red}]{\$\{}{\}},
  postbreak=\mbox{\textcolor{arrowcolor}{$\hookrightarrow$}}
}
\lstdefinestyle{myC2}{
  style=myC,
  keywords=[2]{mapping},
  keywordstyle=[2]\color{cyan},
  emphstyle=\color{deepred}
}

\usepackage{multirow}
\usepackage{pgfplotstable}
\usepackage{pgfplots}
\usepackage{marvosym}
\usepackage[english]{babel}
\usepackage{upgreek}
\usepackage{makecell}
\usepackage{framed}
\usepackage{fp}
\usepackage{setspace}
\usepackage{pifont}
\usepackage[nolist]{acronym}

\def\BibTeX{{\rm B\kern-.05em{\sc i\kern-.025em b}\kern-.08em
    T\kern-.1667em\lower.7ex\hbox{E}\kern-.125emX}}

\usepackage[hidelinks]{hyperref}
\usepackage[backend=biber, style=numeric, hyperref, maxnames=3, minnames=1, giveninits=true, maxcitenames=2, mincitenames=1,bibencoding=utf8, isbn=true,sorting=none]{biblatex}
\usepackage{cleveref}
\usepackage{siunitx}
\usepackage{xspace}
\usepackage{xargs}
\usepackage{todonotes}
\usepackage{csquotes}

\newcommand{\II}{\textit{II}\xspace}

\DeclareSIUnit\flops{FLOPS}
\DeclareSIUnit\flop{FLOP}

\DeclareSIUnit\ops{OPS}
\DeclareSIUnit\op{OP}

\pgfplotsset{compat=1.18}

\graphicspath{{figures/}}

\usepackage{bm}

\DeclareSIUnit\flops{FLOPS}
\DeclareSIUnit\flop{FLOP}

\DeclareSIUnit\ops{OPS}
\DeclareSIUnit\op{OP}

\begin{document}

\newcommand{\tikzpic}[3]{
    \tikzset{pics/#1/.style n args={#2}{code={#3}}}
}
\tikzpic{blockPic}{5}{
    \definecolor{leftColor}{HTML}{ffffff};
    \definecolor{rightColor}{HTML}{f4f4f4};

    \draw[drop shadow={shadow xshift=0.02cm, shadow yshift=-0.02cm, color=black}, shading=axis, left color=leftColor, right color=rightColor, shading angle=0] (0, 0) -- +(#2, 0) -- +(#2, {((#3))}) -- +(0, {((#3))}) -- cycle;
    \node[align=center] at ({((#2)) * 0.5}, {(((#3)) / 2)}) {#1};

    \foreach \numNorth\numEast\numSouth\numWest in {#4} {
        \foreach \north\east\south\west in {#5} {
            \foreach \port [count=\i] in \north {
                \coordinate (\port) at ({((#2)) / (\numNorth + 1) * \i}, #3);
            }
            \foreach \port [count=\i] in \east {
                \coordinate (\port) at (#2, {((#3)) / (\numEast + 1) * \i});
            }
            \foreach \port [count=\i] in \south {
                \coordinate (\port) at ({((#2)) / (\numSouth + 1) * \i}, 0);
            }
            \foreach \port [count=\i] in \west {
                \coordinate (\port) at (0, {((#3)) / (\numWest + 1) * \i});
            }
        }
    }
}
\tikzpic{coloredBlockPic}{7}{

    \foreach \leftColor\rightColor in {#6} {
        \draw[drop shadow={shadow xshift=0.02cm, shadow yshift=-0.02cm, color=black}, shading=axis, left color=\leftColor, right color=\rightColor, shading angle=0] (0, 0) -- +(#2, 0) -- +(#2, {((#3))}) -- +(0, {((#3))}) -- cycle;
        \node[align=center, color=#7] at ({((#2)) * 0.5}, {(((#3)) / 2)}) {#1};

        \foreach \numNorth\numEast\numSouth\numWest in {#4} {
            \foreach \north\east\south\west in {#5} {
                \foreach \port [count=\i] in \north {
                    \coordinate (\port) at ({((#2)) / (\numNorth + 1) * \i}, #3);
                }
                \foreach \port [count=\i] in \east {
                    \coordinate (\port) at (#2, {((#3)) / (\numEast + 1) * \i});
                }
                \foreach \port [count=\i] in \south {
                    \coordinate (\port) at ({((#2)) / (\numSouth + 1) * \i}, 0);
                }
                \foreach \port [count=\i] in \west {
                    \coordinate (\port) at (0, {((#3)) / (\numWest + 1) * \i});
                }
            }
        }
    }
}

\tikzpic{grayBlockPic}{5}{
    \definecolor{leftColor}{HTML}{d9d9d9};
    \definecolor{rightColor}{HTML}{d9d9d9};

    \definecolor{darkPurple}{RGB}{96, 25, 134}   
    \definecolor{darkBlue}{RGB}{0, 59, 111}      
    \definecolor{darkBrown}{RGB}{102, 51, 0}     
    \definecolor{darkGray}{RGB}{71, 79, 82}   
    \definecolor{darkRed}{RGB}{177, 0, 18}
    \definecolor{darkGreen}{RGB}{0, 147, 146}
    \definecolor{darkOrange}{RGB}{255, 111, 0}

    \draw[drop shadow={shadow xshift=0.02cm, shadow yshift=-0.02cm, color=black}, shading=axis, left color=leftColor, right color=rightColor, shading angle=0] (0, 0) -- +(#2, 0) -- +(#2, {((#3))}) -- +(0, {((#3))}) -- cycle;
    \node[align=center] at ({((#2)) * 0.5}, {(((#3)) / 2)}) {#1};

    \foreach \numNorth\numEast\numSouth\numWest in {#4} {
        \foreach \north\east\south\west in {#5} {
            \foreach \port [count=\i] in \north {
                \coordinate (\port) at ({((#2)) / (\numNorth + 1) * \i}, #3);
            }
            \foreach \port [count=\i] in \east {
                \coordinate (\port) at (#2, {((#3)) / (\numEast + 1) * \i});
            }
            \foreach \port [count=\i] in \south {
                \coordinate (\port) at ({((#2)) / (\numSouth + 1) * \i}, 0);
            }
            \foreach \port [count=\i] in \west {
                \coordinate (\port) at (0, {((#3)) / (\numWest + 1) * \i});
            }
        }
    }
}
\tikzpic{emptyBlockPic}{5}{
    \foreach \numNorth\numEast\numSouth\numWest in {#4} {
        \foreach \north\east\south\west in {#5} {
            \foreach \port [count=\i] in \north {
                \coordinate (\port) at ({((#2)) / (\numNorth + 1) * \i}, #3);
            }
            \foreach \port [count=\i] in \east {
                \coordinate (\port) at (#2, {((#3)) / (\numEast + 1) * \i});
            }
            \foreach \port [count=\i] in \south {
                \coordinate (\port) at ({((#2)) / (\numSouth + 1) * \i}, 0);
            }
            \foreach \port [count=\i] in \west {
                \coordinate (\port) at (0, {((#3)) / (\numWest + 1) * \i});
            }
        }
    }
}

\tikzpic{aluPic}{3}{
    \pgfmathsetlengthmacro{\aluUnitSize}{{((#2) / 5)}};
    \definecolor{leftColor}{HTML}{ffffff};
    \definecolor{rightColor}{HTML}{f4f4f4};

    \draw[drop shadow={shadow xshift=0.02cm, shadow yshift=-0.02cm, color=black}, shading=axis, left color=leftColor, right color=rightColor, shading angle=0] 
        (\aluUnitSize, 0) -- 
        (4 * \aluUnitSize, 0) -- 
        (5 * \aluUnitSize, {((#3))}) -- 
        (4 * \aluUnitSize, {((#3))}) -- 
        (3.5 * \aluUnitSize, {((#3) / 2)}) -- 
        (1.5 * \aluUnitSize, {((#3) / 2)}) -- 
        (1 * \aluUnitSize, {((#3))}) -- 
        (0 * \aluUnitSize, {((#3))}) -- 
        (\aluUnitSize, 0);

    \node[align=center] at ({((#2)) * 0.5}, {(((#3)) / 3)}) {#1};

    \coordinate (East0) at (4.25 * \aluUnitSize, {((#3) / 4)});
    \coordinate (East1) at (4.75 * \aluUnitSize, {((#3) / 4) * 3});

    \coordinate (West) at (0.5 * \aluUnitSize, {((#3) / 2)});

    \coordinate (Rs0) at (0.5 * \aluUnitSize, {((#3))});
    \coordinate (Rs1) at (4.5 * \aluUnitSize, {((#3))});

    \coordinate (Rd) at (2.5 * \aluUnitSize, 0) (Rd);
}

\tikzpic{westOpenBlockPic}{5}{
    \definecolor{leftColor}{HTML}{ffffff};
    \definecolor{rightColor}{HTML}{f4f4f4};
    \draw[drop shadow={shadow xshift=0.02cm, shadow yshift=-0.02cm, color=black}, shading=axis, left color=leftColor, right color=rightColor, shading angle=0] (0, 0) -- +(#2, 0) -- +(#2, {((#3))}) -- +(0, {((#3))}) -- cycle;

    \draw[color=white, fill=white] (-0.04cm+#2, -0.04cm) -- +(0.08cm, 0) -- +(0.08cm, {((#3))+0.08cm}) -- +(0, {((#3))+0.08cm}) -- +(0, 0);

    \node[align=center] at ({((#2)) * 0.5}, {(((#3)) / 2)}) {#1};

    \foreach \numNorth\numEast\numSouth\numWest in {#4} {
        \foreach \north\east\south\west in {#5} {
            \foreach \port [count=\i] in \north {
                \coordinate (\port) at ({((#2)) / (\numNorth + 1) * \i}, #3);
            }
            \foreach \port [count=\i] in \east {
                \coordinate (\port) at (#2, {((#3)) / (\numEast + 1) * \i});
            }
            \foreach \port [count=\i] in \south {
                \coordinate (\port) at ({((#2)) / (\numSouth + 1) * \i}, 0);
            }
            \foreach \port [count=\i] in \west {
                \coordinate (\port) at (0, {((#3)) / (\numWest + 1) * \i});
            }
        }
    }
}
\tikzpic{eastWestOpenBlockPic}{5}{
    \definecolor{leftColor}{HTML}{ffffff};
    \definecolor{rightColor}{HTML}{f4f4f4};
    \draw[drop shadow={shadow xshift=0.02cm, shadow yshift=-0.02cm, color=black}, shading=axis, left color=leftColor, right color=rightColor, shading angle=0] (0, 0) -- +(#2, 0) -- +(#2, {((#3))}) -- +(0, {((#3))}) -- cycle;

    \draw[color=white, fill=white] (-0.04cm, -0.04cm) -- +(0.08cm, 0) -- +(0.08cm, {((#3))+0.08cm}) -- +(0, {((#3))+0.08cm}) -- +(0, 0);
    \draw[color=white, fill=white] (-0.04cm+#2, -0.04cm) -- +(0.08cm, 0) -- +(0.08cm, {((#3))+0.08cm}) -- +(0, {((#3))+0.08cm}) -- +(0, 0);

    \node[align=center] at ({((#2)) * 0.5}, {(((#3)) / 2)}) {#1};

    \foreach \numNorth\numEast\numSouth\numWest in {#4} {
        \foreach \north\east\south\west in {#5} {
            \foreach \port [count=\i] in \north {
                \coordinate (\port) at ({((#2)) / (\numNorth + 1) * \i}, #3);
            }
            \foreach \port [count=\i] in \east {
                \coordinate (\port) at (#2, {((#3)) / (\numEast + 1) * \i});
            }
            \foreach \port [count=\i] in \south {
                \coordinate (\port) at ({((#2)) / (\numSouth + 1) * \i}, 0);
            }
            \foreach \port [count=\i] in \west {
                \coordinate (\port) at (0, {((#3)) / (\numWest + 1) * \i});
            }
        }
    }
}

\tikzpic{southOpenBlockPic}{5}{
    \definecolor{leftColor}{HTML}{ffffff};
    \definecolor{rightColor}{HTML}{f4f4f4};
    \draw[drop shadow={shadow xshift=0.02cm, shadow yshift=-0.02cm, color=black}, shading=axis, left color=leftColor, right color=rightColor, shading angle=0] (0, 0) -- +(#2, 0) -- +(#2, {((#3))}) -- +(0, {((#3))}) -- cycle;

    \draw[color=white, fill=white] (-0.04cm, 0.02cm) -- +(#2+0.08cm, 0.02cm) -- +(#2+0.08cm, -0.08cm) -- +(-0.04cm, -0.08cm) -- (0, 0);

    \node[align=center] at ({((#2)) * 0.5}, {(((#3)) / 2)}) {#1};

    \foreach \numNorth\numEast\numSouth\numWest in {#4} {
        \foreach \north\east\south\west in {#5} {
            \foreach \port [count=\i] in \north {
                \coordinate (\port) at ({((#2)) / (\numNorth + 1) * \i}, #3);
            }
            \foreach \port [count=\i] in \east {
                \coordinate (\port) at (#2, {((#3)) / (\numEast + 1) * \i});
            }
            \foreach \port [count=\i] in \south {
                \coordinate (\port) at ({((#2)) / (\numSouth + 1) * \i}, 0);
            }
            \foreach \port [count=\i] in \west {
                \coordinate (\port) at (0, {((#3)) / (\numWest + 1) * \i});
            }
        }
    }
}

\tikzpic{rotatedBlockPic}{5}{
    \definecolor{leftColor}{HTML}{ffffff};
    \definecolor{rightColor}{HTML}{f4f4f4};
    \draw[drop shadow={shadow xshift=0.02cm, shadow yshift=-0.02cm, color=black}, shading=axis, left color=leftColor, right color=rightColor, shading angle=0] (0, 0) -- +(#2, 0) -- +(#2, {((#3))}) -- +(0, {((#3))}) -- cycle;
    \node[align=center, rotate=90] at ({((#2)) * 0.5}, {(((#3)) / 2)}) {#1};

    \foreach \numNorth\numEast\numSouth\numWest in {#4} {
        \foreach \north\east\south\west in {#5} {
            \foreach \port [count=\i] in \north {
                \coordinate (\port) at ({((#2)) / (\numNorth + 1) * \i}, #3);
            }
            \foreach \port [count=\i] in \east {
                \coordinate (\port) at (#2, {((#3)) / (\numEast + 1) * \i});
            }
            \foreach \port [count=\i] in \south {
                \coordinate (\port) at ({((#2)) / (\numSouth + 1) * \i}, 0);
            }
            \foreach \port [count=\i] in \west {
                \coordinate (\port) at (0, {((#3)) / (\numWest + 1) * \i});
            }
        }
    }
}

\tikzset{blockArrowTo/.style={-{Latex[length=3pt]}}}
\tikzset{blockArrowFrom/.style={{Latex[length=3pt]}-}}
\tikzset{blockArrowBi/.style={{Latex[length=3pt]}-{Latex[length=3pt]}}}

\tikzpic{peArray}{2}{
    \foreach \x in {1,...,#1} {
        \foreach \y in {1,...,#2} {
            \pic (a) at (\x * 0.65cm - 0.65cm, \y * 0.65cm - 0.65cm) {blockPic={\small PE}{0.5cm}{0.5cm}{2/2/2/2}{{Nin,Nout}/{Ein,Eout}/{Sin,Sout}/{Win, Wout}/}};

            \ifnum\y=#2
                \draw[-] (aNin) -- +(0,  0.6cm);
                \draw[-] (aNout) -- +(0,  0.6cm);
            \else
                \draw[-] (aNin) -- +(0,  0.15cm);
                \draw[-] (aNout) -- +(0,  0.15cm);
            \fi

            \ifnum\x=#1
                \draw[-] (aEin) -- +(0.6cm, 0);
                \draw[-] (aEout) -- +(0.6cm, 0);
            \else
                \draw[-] (aEin) -- +(0.15cm, 0);
                \draw[-] (aEout) -- +(0.15cm, 0);
            \fi

            \ifnum\y=1
                \draw[-] (aSin) -- +(0,  -0.6cm);
                \draw[-] (aSout) -- +(0,  -0.6cm);
            \else
                \draw[-] (aSin) -- +(0,  -0.15cm);
                \draw[-] (aSout) -- +(0,  -0.15cm);
            \fi

            \ifnum\x=1
                \draw[-] (aWin) -- +(-0.60cm, 0);
                \draw[-] (aWout) -- +(-0.60cm, 0);
            \else
                \draw[-] (aWin) -- +(-0.15cm, 0);
                \draw[-] (aWout) -- +(-0.15cm, 0);
            \fi
        }
    }
}

\tikzpic{peArrayNoBuffer}{2}{
    \foreach \x in {1,...,#1} {
        \foreach \y in {1,...,#2} {
            \ifthenelse{\equal{\x}{#1}} {
                \ifthenelse{\equal{\y}{1}} {
                    \node[anchor=south west, minimum width = 0.5cm, minimum height=0.5cm, inner sep = 0] (a) at (\x * 0.65cm - 0.65cm, \y * 0.65cm - 0.65cm) {\textbf{$\ddots$}};
                }{
                    \node[anchor=south west, minimum width = 0.5cm, minimum height=0.5cm, inner sep = 0] (a) at (\x * 0.65cm - 0.65cm, \y * 0.65cm - 0.65cm) {$\dots$};
                }
            }{
                \ifthenelse{\equal{\y}{1}} {
                    \node[anchor=south west, minimum width = 0.5cm, minimum height=0.5cm, inner sep = 0] (a) at (\x * 0.65cm - 0.65cm, \y * 0.65cm - 0.65cm) {$\vdots$};
                }{
                    \pic (a) at (\x * 0.65cm - 0.65cm, \y * 0.65cm - 0.65cm) {blockPic={\footnotesize PE}{0.5cm}{0.5cm}{2/2/2/2}{{Nin,Nout}/{Ein,Eout}/{Sin,Sout}/{Win, Wout}/}};
                    \draw[-] (aNin) -- +(0,  0.15cm);
                    \draw[-] (aNout) -- +(0,  0.15cm);
                    \draw[-] (aEin) -- +(0.15cm, 0);
                    \draw[-] (aEout) -- +(0.15cm, 0);
                    \draw[-] (aSin) -- +(0,  -0.15cm);
                    \draw[-] (aSout) -- +(0,  -0.15cm);
                    \draw[-] (aWin) -- +(-0.15cm, 0);
                    \draw[-] (aWout) -- +(-0.15cm, 0);
                }
            }

        }
    }
}

\tikzpic{baseArray}{2}{
    \pic (Ports) at (0, 0) {grayBlockPic={}{#1*0.65cm+5*0.65cm}{#2*0.65cm+5*0.65cm}{///5}{///{CtrlIn, ConfigIn, DataIn, DataOut}}};
    \pic (array) at (0.65cm*2.5 + 0.15cm/2, 0.65cm*2.5+0.15cm/2) {peArray=
        {#1}
        {#2}
    };

    \node[inner sep=0, anchor=north] at (0.65*2.5+0.15/2, 0.65*2.5+0.15) (SouthWestPeBorder) {};
    \node[inner sep=0, anchor=north] at (0.65*2.5+0.15 + #1 * 0.65 - 0.15, 0.65*2.5+0.15) (SouthEastPeBorder) {};

    \node[inner sep=0, anchor=north] at (0.65*2.5+0.15, 0.65*2.5+0.15+0.65*#2-0.15) (NorthWestPeBorder) {};
    \node[inner sep=0, anchor=north] at (0.65*2.5+0.15 + #1 * 0.65 - 0.15, 0.65*2.5+0.15+0.65*#2-0.15) (NorthEastPeBorder) {};

    \pic (bufferNorth) at (0.65*2.5+0.15/2, 0.65*2.5+0.15/2+#2*0.65+0.45) {blockPic={I/O Buffers}{0.65cm * #1-0.15cm}{1*0.85cm}{}{}};
    \pic (agNorth) at (0.65*2.5+0.15/2, 0.65*2.5+0.15/2+#2*0.65) {blockPic={\small Address Generators}{0.65cm * #1-0.15cm}{1*0.35cm}{}{}};

    \pic (bufferWest) at (0.65*2.5+0.15/2+0.65*#1+0.45, 0.65*2.5+0.15/2) {rotatedBlockPic={I/O Buffers}{1*0.85cm}{0.65cm * #2-0.15cm}{}{}};
    \pic (agWest) at (0.65*2.5+0.15/2+0.65*#1, 0.65*2.5+0.15/2) {rotatedBlockPic={\small Address Generators}{1*0.35cm}{0.65cm * #2-0.15cm}{}{}};

    \pic (bufferSouth) at (0.65*2.5+0.15/2, 0.65*0.5 - 0.15/2) {blockPic={I/O Buffers}{0.65cm * #1-0.15cm}{1*0.85cm}{}{}};
    \pic (agSouth) at (0.65*2.5+0.15/2, 0.65*0.5+0.95 - 0.15/2) {blockPic={\small Address Generators}{0.65cm * #1-0.15cm}{1*0.35cm}{}{}};

    \pic (bufferEast) at (0.65*0.5-0.15/2, 0.65*2.5+0.15/2) {rotatedBlockPic={I/O Buffers}{1*0.85cm}{0.65cm * #2-0.15cm}{}{}};
    \pic (agEast) at (0.65*0.5+0.95-0.15/2, 0.65*2.5+0.15/2) {rotatedBlockPic={\small Address Generators}{1*0.35cm}{0.65cm * #2-0.15cm}{}{}};
}

\tikzpic{tcpaPeriphery}{2}{
    \pic (box) at (0, 0) {grayBlockPic={}{#1}{#2}{///}{///}};

    \pgfmathsetlengthmacro{\innerPadding}{1.0cm};
    \pgfmathsetlengthmacro{\outerPadding}{0.25cm};

    \pgfmathsetlengthmacro{\width}{#1 - \outerPadding - \outerPadding}
    \pgfmathsetlengthmacro{\height}{(#2 - 2 * \outerPadding - 3 * \innerPadding) / 4};

    \pic (Gc) at (\outerPadding, \outerPadding + 3 * \innerPadding + 3 * \height)   {blockPic={\glsentrylong{gc}}{\width}{\height}{/1/1/}{/CtrlOut/ConfigIn/}};
    \pic (Cm) at (\outerPadding, \outerPadding + 2 * \innerPadding + 2 * \height)   {blockPic={Configuration\\Manager}{\width}{\height}{1/1/1/1}{ConfigGC/ConfigPE/ConfigLion/ConfigIn}};
    \pic (Lion) at (\outerPadding, \outerPadding + 1 * \innerPadding + 1 * \height) {blockPic={\glsentrylong{lion}}{\width}{\height}{1//1/1}{ConfigIn//DMAOut/RequestsOut}};
    \pic (Bif) at (\outerPadding, \outerPadding)                                    {blockPic={DMA}{\width}{\height}{1/2/1/2}{DMAIn/{DataArrayIn, DataArrayOut}/DebugIn/{DataCtrlIn, DataCtrlOut}}};
} 

\tikzpic{alpacaFPGA}{2}{

    \pgfmathsetlengthmacro{\innerPadding}{1.0cm};
    \pgfmathsetlengthmacro{\outerPadding}{0.25cm};
    \pgfmathsetlengthmacro{\labelHeight}{0.5cm};

    \pgfmathsetlengthmacro{\width}{#1 - \outerPadding - \outerPadding}
    \pgfmathsetlengthmacro{\height}{(#2 - \labelHeight - 2 * \outerPadding - 3 * \innerPadding) / 4};

    \pic (box) at (0, 0) {grayBlockPic={}{#1}{#2}{///}{///}};
    \pic (box) at (0, 0) {blockPic={Host}{#1}{\labelHeight}{///}{///}};

    \pic (Clock) at (\outerPadding, \outerPadding + 3 * \innerPadding + 3 * \height + \labelHeight)   {blockPic={Clock}{\width}{\height}{/1//}{/Out//}};
    \pic (CPU) at (\outerPadding, \outerPadding + 2 * \innerPadding + 2 * \height + \labelHeight)   {blockPic={CPU}{\width}{\height}{//1/}{//Out/}};
    \pic (DMA) at (\outerPadding, \outerPadding + 1 * \innerPadding + 1 * \height + \labelHeight) {blockPic={DMA}{\width}{\height}{1/2/2/}{CtrlIn/{ExtOut, ExtIn}/{IntOut, IntIn}/}};
    \pic (Memory) at (\outerPadding, \outerPadding + \labelHeight)                                    {blockPic={MEM}{\width}{\height}{2///}{{In, Out}///}};
} 

\tikzpic{alpaca}{0}{
    \pic (fpga) at (0, 0) {alpacaFPGA={2.5cm}{8*0.65cm+5*0.65cm}};

    \pic (tcpa) at (8.5, 0) {baseArray={8}{8}};
    \pic (ctrl) at (3, 0) {rotatedBlockPic={I/O}{0.5cm}{8*0.65cm+5*0.65cm}{0/5/0/1}{/{ConfigOut, RequestsIn, DataIn, DataOut, DebugIn}/{}/{FPGA}}};
    \pic (peri) at (4, 0) {tcpaPeriphery={4cm}{8*0.65cm+5*0.65cm}};

    \draw[blockArrowTo]   (periGcCtrlOut.east)  -- (periGcCtrlOut.east -| tcpaPortsCtrlIn.west);
    \draw[blockArrowTo]   (periCmConfigGC)      -- (periCmConfigGC |- periGcConfigIn);
    \draw[blockArrowTo]   (periCmConfigPE)      -- (periCmConfigPE -| tcpaPortsConfigIn);
    \draw[blockArrowTo]   (periCmConfigLion)    -- (periCmConfigLion |- periLionConfigIn);
    \draw[blockArrowFrom] (periCmConfigIn)      -- (periCmConfigIn -| ctrlConfigOut);
    \draw[blockArrowTo]   (periLionRequestsOut) -- (periLionRequestsOut -| ctrlRequestsIn);
    \draw[blockArrowTo]   (periLionDMAOut)      -- (periBifDMAIn);
    \draw[blockArrowFrom] (periBifDataCtrlIn)   -- (periBifDataCtrlIn -| ctrlDataIn);
    \draw[blockArrowTo]   (periBifDataCtrlOut)  -- (periBifDataCtrlOut -| ctrlDataOut);
    \draw[blockArrowTo]   (periBifDataArrayIn)  -- (periBifDataArrayIn -| tcpaPortsDataIn);
    \draw[blockArrowFrom] (periBifDataArrayOut) -- (periBifDataArrayOut -| tcpaPortsDataOut);

    \draw[blockArrowTo]   (fpgaClockOut)  -- (fpgaClockOut -| ctrlFPGA);

    \draw[blockArrowTo]   (fpgaCPUOut)  -- (fpgaCPUOut |- fpgaDMACtrlIn);

    \draw[blockArrowTo]   (fpgaDMAExtOut)  -- (fpgaDMAExtOut -| ctrlFPGA);
    \draw[blockArrowFrom]   (fpgaDMAExtIn)  -- (fpgaDMAExtIn -| ctrlFPGA);

    \draw[blockArrowTo]   (fpgaDMAIntOut)  -- (fpgaMemoryIn);
    \draw[blockArrowFrom]   (fpgaDMAIntIn)  -- (fpgaMemoryOut);
}

\tikzpic{bufferModuleWithoutBus}{2}{
    \foreach \x in {1,...,#1} {
        \pic (bank) at ({1.5cm * (\x - 1)}, 4.5) {rotatedBlockPic={Bank}{1cm}{3cm}{2//2/}{{Nin,Nout}//{Sin,Sout}/}};
        \draw[blockArrowFrom] (bankNin) -- +(0, 0.5cm);
        \draw[blockArrowTo] (bankNout) -- +(0, 0.5cm);

        \draw[blockArrowFrom] (bankSin) -- +(0, -0.5cm);
        \draw[blockArrowTo] (bankSout) -- +(0, -0.5cm);
    }

    \pic (mux) at (0, 3.5cm) {blockPic={Data and Address Crossbar}{#1 * 1.5cm}{0.5cm}{//1/}{//In/}};

    \pgfmathsetlengthmacro{\agSizeWithPadding}{(#1 * 1.5cm) / #2};
    \pgfmathsetlengthmacro{\agPadding}{1.5cm - \agSizeWithPadding};

    \foreach \x in {1,...,#2} {
        \pic (ag) at ({\agSizeWithPadding * (\x - 1)}, 1.5cm) {blockPic={AG}{1.5cm}{1.5cm}{1//2/}{Nout//{Sin,Sout}/}};
        \draw[blockArrowTo] let \p1 = (agNout), \p2=(muxIn) in (agNout) -- (\x1, \y2);

        \draw[blockArrowTo] ([xshift=1.5cm, yshift=-0.5cm] agSout.north) -- ([xshift=1.5cm] agSout.north |- muxIn);
        \draw[blockArrowFrom] ([xshift=1.5cm, yshift=-0.5cm] agSin.north) -- ([xshift=1.5cm] agSin.north |- muxIn);
    }

    \pic (pe) at (0, 0) {southOpenBlockPic={PE}{#1 * 1.5cm}{1.0cm}{}{}};
}

\tikzpic{lion}{2}{
    \pic (box) at (0, 0) {grayBlockPic={}{#1}{#2}{///}{///}};

    \pgfmathsetlengthmacro{\innerPadding}{0.5cm};
    \pgfmathsetlengthmacro{\outerPadding}{0.25cm};
    \pgfmathsetlengthmacro{\labelHeight}{0.5cm};
    \pgfmathsetlengthmacro{\rightOffset}{\innerPadding + \innerPadding};

    \pgfmathsetlengthmacro{\width}{#1 - \outerPadding - \rightOffset}
    \pgfmathsetlengthmacro{\height}{(#2 - 2 * \outerPadding - 2 * \innerPadding - \labelHeight) / 3};

    \pic (pq) at (\outerPadding, \labelHeight + \outerPadding) {blockPic={Priority\\Queue}{\width}{\height}{1/1//}{Out/In//}};
    \pic (config) at (\outerPadding, \labelHeight + \outerPadding + \height + \innerPadding) {blockPic={Configuration\\Memory}{\width}{\height}{1/1/1/}{Out/In/Addr/}};
    \pic (Dma) at (\outerPadding, \labelHeight + \outerPadding + \height + \innerPadding + \height + \innerPadding) {blockPic={Address\,and\,Deadline\\Computation}{\width}{\height}{1/2/1/1}{RequestToDMA/{Config, Addr}/In/RequestToIO}};
    
    \draw[blockArrowTo] (pqOut) -- (configAddr);
    \draw[blockArrowTo] (configOut) -- (DmaIn);
    \draw[blockArrowTo] (DmaConfig) -- +(\rightOffset / 3 * 1, 0) |- (configIn);
    \draw[blockArrowTo] (DmaAddr) -- +(\rightOffset / 3 * 2, 0) |- (pqIn);

    \pic at (0, 0) {blockPic={\glsentrylong{lion}}{#1}{\labelHeight}{}{}};
} 

\tikzpic{bufferDMA}{2}{
    \pic (box) at (0, 0) {grayBlockPic={}{#1}{#2}{///}{///}};

    \pgfmathsetlengthmacro{\innerPadding}{0.5cm};
    \pgfmathsetlengthmacro{\outerPadding}{0.25cm};
    \pgfmathsetlengthmacro{\busHeight}{0.5cm};
    \pgfmathsetlengthmacro{\busOverlap}{0.3cm};

    \pgfmathsetlengthmacro{\width}{#1 - \outerPadding - \outerPadding}
    \pgfmathsetlengthmacro{\height}{(#2 - 2 * \outerPadding - 1 * \innerPadding - \busHeight) / 1};

    \pic (Dma) at (\outerPadding, \outerPadding) {blockPic={DMA}{\width}{\height}{2//1/2}{{BusIn, BusOut}//Request/{IOin, IOout}}};
    \pic (bus) at (-\busOverlap, #2 - \busHeight - \outerPadding) {eastWestOpenBlockPic={Buffer Bus}{#1 + 2 * \busOverlap}{\busHeight}{//2/}{//{In, Out}/}};
    
    \draw[blockArrowTo] (DmaBusOut) -- (DmaBusOut |- busIn);
    \draw[blockArrowFrom] (DmaBusIn) -- (DmaBusIn |- busOut);
}

\tikzpic{bufferModule}{3}{
    \pic (box) at (7cm, -3.25cm+4.5cm+2cm+0.125cm) {grayBlockPic={}{#3 * #1 * 1.5cm + #3 * 0.5cm}{3.25cm+3.5cm+0.75cm-2cm-0.125cm}{///}{///}};
    \pic (box) at (7cm, -3.25cm+4.5cm) {grayBlockPic={}{#3 * #1 * 1.5cm + #3 * 0.5cm}{1.875cm}{///}{///}};
    \pic (bus) at (6.7cm, 3.5cm+4.5cm) {eastWestOpenBlockPic={Buffer Bus}{#3 * #1 * 1.5cm + #3 * 0.5cm + 0.6cm}{0.5cm}{}{}};
    \foreach \p in {1,...,#3} {
        \pic (module) at ({7.25cm + (#1 * 1.5cm + 0.5cm) * (\p - 1)}, 0) {bufferModuleWithoutBus={#1}{#2}};
    }

    \pic (lion) at (1cm, -3.25cm+4.5cm) {lion={5cm}{4.75cm}};

    \pic (dma) at (1cm, 2cm+4.5cm) {bufferDMA={5cm}{2.25cm}};

    \draw[blockArrowTo] (lionDmaRequestToDMA) -- (lionDmaRequestToDMA |- dmaDmaRequest);
    \draw[blockArrowTo] (lionDmaRequestToIO) -- +(-1cm, 0);

    \draw[blockArrowTo] (dmaDmaIOout) -- +(-1cm, 0);
    \draw[blockArrowFrom] (dmaDmaIOin) -- +(-1cm, 0);

}

\tikzpic{globalController}{7}{
    \pic (box) at (0, 0) {grayBlockPic={}{#1}{#2}{///}{///}};

    \pgfmathsetlengthmacro{\innerPadding}{0.75cm};
    \pgfmathsetlengthmacro{\outerPadding}{0.25cm};
    \pgfmathsetlengthmacro{\labelHeight}{0.5cm};

    \pgfmathsetlengthmacro{\repeatOffset}{\innerPadding * 0.75 / (max(#3, #4, #5, #6, #7))};
    \pgfmathsetlengthmacro{\repeatBottomPadding}{\repeatOffset * (#3 - 1) + \labelHeight};
    \pgfmathsetlengthmacro{\repeatBottomFullPadding}{\repeatOffset * (max(#6, #7) - 1) + \labelHeight};

    \pgfmathsetlengthmacro{\subwidth}{(#1 - 2 * \outerPadding - 3 * \innerPadding - \repeatOffset * (#7 - 1)) / 8};
    \pgfmathsetlengthmacro{\width}{\subwidth * 3};

    \pgfmathsetlengthmacro{\height}{(#2 - 2 * \outerPadding - 2 * \innerPadding - \repeatBottomPadding) / 3};
    \pgfmathsetlengthmacro{\leftInnerPadding}{(#2 - 2 * \height - \labelHeight) / 3};

    \pgfmathsetlengthmacro{\fullWidth}{\subwidth};
    \pgfmathsetlengthmacro{\fullHeight}{#2 - 2 * \outerPadding - \repeatBottomFullPadding};

    \pic (pulse) at ({\outerPadding + 0 * (\width + \innerPadding)}, {\labelHeight + 2 * \leftInnerPadding + \height}) {blockPic={Pulse\\Generator}{\width}{\height}{//1/}{//Out/}};
    \pic (scanner) at ({\outerPadding + 0 * (\width + \innerPadding)}, {\labelHeight + \leftInnerPadding}) {blockPic={Iteration\,Space\\Scanner}{\width}{\height}{1/1//}{In/Out//}};

    \foreach \i in {#7,...,1} {
        \pic (orsel) at ({\outerPadding + 2 * (\width + \innerPadding) + \fullWidth + \innerPadding + (\i - 1) * \repeatOffset}, {\outerPadding + 0 * (\height + \innerPadding) + \repeatBottomFullPadding - (\i - 1) * \repeatOffset}) {rotatedBlockPic={Logic Linker}{\fullWidth}{\fullHeight}{/1//1}{/Out//In}};
        \draw[blockArrowTo] (orselOut) -- +({\repeatOffset * (#7 - 1) + 2 * \outerPadding - (\i - 1) * \repeatOffset}, 0);
    }

    \foreach \i in {#3,...,1} {
        \pic (ineq) at ({\outerPadding + 1.25 * (\width + \innerPadding) + (\i - 1) * \repeatOffset}, {\outerPadding + 0 * (\height + \innerPadding) + \repeatBottomPadding - (\i - 1) * \repeatOffset}) {blockPic={Inequation\\Comparator}{\width}{\height}{/1//1}{/Out//In}};
        \draw[blockArrowTo] (scannerOut) -- +(0.5 * \innerPadding, 0) |- (ineqIn);
        \draw[blockArrowTo] (ineqOut) -- (ineqOut -| orselIn);
    }
    \foreach \i in {#4,...,1} {
        \pic (lower) at ({\outerPadding + 1.25 * (\width + \innerPadding) + (\i - 1) * \repeatOffset}, {\outerPadding + 1 * (\height + \innerPadding) + \repeatBottomPadding - (\i - 1) * \repeatOffset}) {blockPic={Lower Bound\\Comparator}{\width}{\height}{/1//1}{/Out//In}};
        \draw[blockArrowTo] (scannerOut) -- +(0.5 * \innerPadding, 0) |- (lowerIn);
        \draw[blockArrowTo] (lowerOut) -- (lowerOut -| orselIn);
    }
    \foreach \i in {#5,...,1} {
        \pic (upper) at ({\outerPadding + 1.25 * (\width + \innerPadding) + (\i - 1) * \repeatOffset}, {\outerPadding + 2 * (\height + \innerPadding) + \repeatBottomPadding - (\i - 1) * \repeatOffset}) {blockPic={Upper Bound\\Comparator}{\width}{\height}{/1//1}{/Out//In}};
        \draw[blockArrowTo] (scannerOut) -- +(0.5 * \innerPadding, 0) |- (upperIn);
        \draw[blockArrowTo] (upperOut) -- (upperOut -| orselIn);
    }

    \draw[blockArrowTo] (pulseOut) -- (scannerIn);
    \pic at (0, 0) {blockPic={\glsentrylong{gc}}{#1}{\labelHeight}{}{}};
}

\tikzpic{addressGenerator}{2}{
    \pic (box) at (0, 0) {grayBlockPic={}{#1}{#2}{///}{///}};

    \pgfmathsetlengthmacro{\innerPadding}{0.5cm};
    \pgfmathsetlengthmacro{\outerPadding}{0.25cm};
    \pgfmathsetlengthmacro{\labelHeight}{0.5cm};

    \pgfmathsetlengthmacro{\width}{(#1 - 2 * \outerPadding - 2 * \innerPadding) / 3};
    \pgfmathsetlengthmacro{\height}{(#2 - 2 * \outerPadding - 2 * \innerPadding - \labelHeight) / 3};

    \pic (pulse) at      ({\outerPadding + 0 * (\width + \innerPadding)}, {\labelHeight + \outerPadding + 2 * (\height + \innerPadding)}) {blockPic={Pulse\\Generator}{\width}{\height}{/1//}{/Out//}};
    \pic (scanner) at    ({\outerPadding + 1 * (\width + \innerPadding)}, {\labelHeight + \outerPadding + 2 * (\height + \innerPadding)}) {blockPic={Iteration Space\\Scanner}{\width}{\height}{/1/1/1}{/ToBox/ToIneq/In}};
    \pic (bounding) at   ({\outerPadding + 2 * (\width + \innerPadding)}, {\labelHeight + \outerPadding + 2 * (\height + \innerPadding)}) {blockPic={Boundingbox\\Comparator}{\width}{\height}{//1/1}{//Out/In}};
    \pic (ineq) at       ({\outerPadding + 1 * (\width + \innerPadding)}, {\labelHeight + \outerPadding + 1 * (\height + \innerPadding)}) {blockPic={Inequation\\Comparator}{\width}{\height}{1//1/}{In//Out/}};
    \pic (subscanner) at ({\outerPadding + 2 * (\width + \innerPadding)}, {\labelHeight + \outerPadding + 1 * (\height + \innerPadding)}) {blockPic={Iteration Space\\Scanner}{\width}{\height}{1//1/}{In//Out/}};
    \pic (address) at    ({\outerPadding + 2 * (\width + \innerPadding)}, {\labelHeight + \outerPadding + 0 * (\height + \innerPadding)}) {blockPic={Address\\Computation}{\width}{\height}{1///1}{In///Out}};
    \pic (banksel) at    ({\outerPadding + 1 * (\width + \innerPadding)}, {\labelHeight + \outerPadding + 0 * (\height + \innerPadding)}) {blockPic={Bank\\Selector}{\width}{\height}{1/1//1}{EnIn/AddrIn//Out}};
    \pic (abuf) at       ({\outerPadding + 0 * (\width + \innerPadding)}, {\labelHeight + \outerPadding + 0 * (\height + \innerPadding)}) {blockPic={Address\\Buffer}{\width}{\height}{/1//1}{/In//Out}};

    \draw[blockArrowTo] (pulseOut) -- (scannerIn);
    \draw[blockArrowTo] (scannerToBox) -- (boundingIn);
    \draw[blockArrowTo] (scannerToIneq) -- (ineqIn);
    \draw[blockArrowTo] (boundingOut) -- (subscannerIn);
    \draw[blockArrowTo] (subscannerOut) -- (addressIn);
    \draw[blockArrowTo] (addressOut) -- (bankselAddrIn);
    \draw[blockArrowTo] (ineqOut) -- (bankselEnIn);
    \draw[blockArrowTo] (bankselOut) -- (abufIn);
    \draw[blockArrowTo] (abufOut) -- +(-2 * \outerPadding, 0);

    \pic at (0, 0) {blockPic={Address Generator}{#1}{\labelHeight}{}{}};
}

\tikzpic{functionalUnit}{1}{
    \pic (box) at (0, 0cm) {grayBlockPic={}{6cm}{2.5cm}{}{}};
    \pic (virtualTarget) at (-0.75cm, 0) {emptyBlockPic={}{0.5cm}{0.5cm}{/1//}{/In//}};

    \pic (virtualCtrlTarget) at (6.25cm, 0) {emptyBlockPic={}{0.5cm}{0.5cm}{///1}{///In}};

    \pic (decoder) at (3cm, 0.125cm) {rotatedBlockPic={Decoder}{0.5cm}{2.125cm}{/2//1}{/{MemIn,BranchOut}//Opcode}};
    \pic (decoderRegs) at (3cm, 0.125cm) {emptyBlockPic={Decoder}{0.5cm}{0.5cm}{///3}{///{Rs0,Rs1,Rd}}};

    \pic (branch) at (0.125cm + 3.75cm, 1.750cm) {blockPic={Branch}{2.0cm}{0.5cm}{/2/1/1}{/{ToCtrl,FromCtrl}/Out/In}};
    \pic (mem) at (0.125cm + 3.75cm, 0.125cm) {blockPic={Instruction\\Memory}{2.0cm}{1.375cm}{1///1}{In///Out}};

    \pic (alu) at (0.25cm, 0.75cm) {aluPic={#1}{2.5cm}{1.25cm}{}{}};

    \draw[blockArrowTo] (branchOut) -- (branchOut |- memIn);
    \draw[blockArrowTo] (memOut) -- (memOut -| decoderMemIn);
    \draw[blockArrowFrom] (branchIn) -- (branchIn -| decoderBranchOut);

    \draw[blockArrowTo] (branchToCtrl) -- (branchToCtrl -| virtualCtrlTargetIn);
    \draw[blockArrowFrom] (branchFromCtrl) -- (branchFromCtrl -| virtualCtrlTargetIn);

    \draw[blockArrowFrom] (aluEast0) -- (aluEast0 -| decoderOpcode);

    \draw[blockArrowTo] (aluEast1) -- +(0.25cm, 0) -- +(0.25cm, 0.70cm) -- ([yshift=0.7cm]aluEast1 -| virtualCtrlTargetIn);

    \draw[blockArrowTo] (decoderRegsRs0) -- (decoderRegsRs0 -| virtualTargetIn);
    \draw[blockArrowTo] (decoderRegsRs1) -- (decoderRegsRs1 -| virtualTargetIn);
    \draw[blockArrowTo] (decoderRegsRd) --  (decoderRegsRd -| virtualTargetIn);

    \draw[blockArrowTo] (aluRd) -- +(0, -0.1cm) -- ([yshift=-0.1cm]aluRd -| virtualTargetIn);
    \draw[blockArrowFrom] (aluRs0) -- +(0, 0.3cm) -- ([yshift=0.3cm]aluRs0 -| virtualTargetIn);
    \draw[blockArrowFrom] (aluRs1) -- +(0, 0.4cm) -- ([yshift=0.4cm]aluRs1 -| virtualTargetIn);
}

\tikzpic{fifo}{3}{
    \foreach \p in {#3,...,1} {
        \pic at (0, {(\p - 1) * #2}) {blockPic={}{#1}{#2}{}{}};
    }
}

\tikzpic{fifoRegisters}{5}{
    \pic at (0, 0) {blockPic={}{#1}{#2}{/2//2}{/{RightIn, RightOut}//{LeftOut, LeftIn}}};

    \pgfmathsetlengthmacro{\horizontalPadding}{((((#1)) - #3 * #4) / (#3 + 1))};
    \pgfmathsetmacro{\slotCount}{scalar((((#2)) - 0.25cm - 0.5cm) / 0.15cm)};

    \foreach \p in {1,...,#3} {
        \pic at (\horizontalPadding * \p + #4 * \p - #4, 0.5cm + 0.125cm) {fifo={#4}{0.15cm}{\slotCount}};
        \coordinate (Id\p) at (\horizontalPadding * \p + #4 * \p - #4 + #4 / 2, #2);
        \coordinate (Od\p) at (\horizontalPadding * \p + #4 * \p - #4 + #4 / 2, 0);
    }

    \pic at (0, 0) {blockPic={#5}{#1}{0.5cm}{}{}};
}

\tikzpic{registerFile}{1}{
    \pgfmathsetlengthmacro{\totalHeight}{#1 * (2.5cm + 0.125cm) - 0.125cm};
    \pgfmathsetlengthmacro{\padding}{(\totalHeight - 1.0cm - 1.0cm - 4.5cm - 3.0cm - 0.125cm * 2) / 3};

    \pic at (-0.125cm, -0.125cm) {grayBlockPic={}{7.50cm}{\totalHeight}{}{}};

    \pic (regFile) at (6.625cm+0.25cm, -0.125cm) {rotatedBlockPic={Data Register File}{0.5cm}{\totalHeight}{///2}{///{In, Out}}};

    \pic (ods) at (0, 0cm) {blockPic={Output Registers}{6.5cm}{1.0cm}{/2/6/}{/{In, Out}/{Od0, Od1, Od2, Od3, Od4, Od5}/}};
    \pic (dataS) at (0, -1cm) {blockPic={Data Interconnect}{6.5cm}{0.5cm}{6///}{{Id0, Id1, Id2, Id3, Id4, Id5}///}};

    \pic (fds) at (0.0cm, 1.0cm + \padding) {fifoRegisters={6.5cm}{4.5cm}{3}{1cm}{Feedback Registers}};

    \pic (rds) at (0, 1.0cm + 4.5cm + 2 * \padding) {blockPic={General Purpose Registers}{6.5cm}{1.0cm}{/2//}{/{In, Out}//}};

    \pic (ids) at (0.0cm, 1.0cm + 4.5cm + 1.0cm + 3 * \padding) {fifoRegisters={6.5cm}{3.0cm}{6}{1cm}{Input Registers}};
    \pic (dataN) at (0.0cm, 1.0cm + 4.5cm + 1.0cm + 3 * \padding + 3.5cm) {blockPic={Data Interconnect}{6.5cm}{0.5cm}{//6/}{//{Od0, Od1, Od2, Od3, Od4, Od5}/}};

    \draw[blockArrowTo] (odsOut) -- (odsOut -| regFileIn);
    \draw[blockArrowFrom] (odsIn) -- (odsIn -| regFileOut);

    \draw[blockArrowTo] (rdsOut) -- (rdsOut -| regFileIn);
    \draw[blockArrowFrom] (rdsIn) -- (rdsIn -| regFileOut);

    \draw[blockArrowTo] (idsRightOut) -- (idsRightOut -| regFileIn);
    \draw[blockArrowFrom] (idsRightIn) -- (idsRightIn -| regFileOut);

    \draw[blockArrowTo] (fdsRightOut) -- (fdsRightOut -| regFileIn);
    \draw[blockArrowFrom] (fdsRightIn) -- (fdsRightIn -| regFileOut);

    \draw[blockArrowTo] (odsOd0) -- (dataSId0);
    \draw[blockArrowTo] (odsOd1) -- (dataSId1);
    \draw[blockArrowTo] (odsOd2) -- (dataSId2);
    \draw[blockArrowTo] (odsOd3) -- (dataSId3);
    \draw[blockArrowTo] (odsOd4) -- (dataSId4);
    \draw[blockArrowTo] (odsOd5) -- (dataSId5);

    \draw[blockArrowFrom] (idsId1) -- (idsId1 |- dataNOd0);
    \draw[blockArrowFrom] (idsId2) -- (idsId2 |- dataNOd1);
    \draw[blockArrowFrom] (idsId3) -- (idsId3 |- dataNOd2);
    \draw[blockArrowFrom] (idsId4) -- (idsId4 |- dataNOd3);
    \draw[blockArrowFrom] (idsId5) -- (idsId5 |- dataNOd4);
}

\tikzpic{ctrlRegisterFile}{1}{
    \pgfmathsetlengthmacro{\totalHeight}{#1 * (2.5cm + 0.125cm) - 0.125cm};
    \pgfmathsetlengthmacro{\padding}{(\totalHeight - 1.0cm - 1.0cm - 7.75cm - 0.125cm * 2) / 2};

    \pic at (-0.625cm-0.25cm, -0.125cm) {grayBlockPic={}{3.750cm+0.75cm}{\totalHeight}{}{}};
    \pic (regFile) at (-0.625cm-0.25cm, -0.125cm) {rotatedBlockPic={Control Register File}{0.5cm}{\totalHeight}{/2//}{/{In, Out}//}};

    \pic (ods) at (0, 0cm) {blockPic={Output Registers}{3.5cm}{1.0cm}{//7/2}{//{Od0, Od1, Od2, Od3, Od4, Od5, Od6}/{In, Out}}};
    \pic  (ctrlS) at (0, -1cm) {blockPic={Control\,Interconnect}{3.5cm}{0.5cm}{7///}{{Id0, Id1, Id2, Id3, Id4, Id5, Id6}///}};
    \pic (flags) at (0, 1.0cm + \padding) {blockPic={Functional Unit\\Flags}{3.5cm}{1.0cm}{///2}{///{In, Out}}};
    \pic (ids) at (0.0cm, 1.0cm + 1.0cm + 2 * \padding) {fifoRegisters={3.5cm}{7.75cm}{7}{0.25cm}{Input Registers}};
    \pic  (ctrlN) at (0.0cm, 1.0cm + 1.0cm + 2 * \padding + 8.25cm) {blockPic={Control\,Interconnect}{3.5cm}{0.5cm}{//7/}{//{Od0, Od1, Od2, Od3, Od4, Od5, Od6}/}};

    \draw[blockArrowTo] (odsOut) -- (odsOut -| regFileIn);
    \draw[blockArrowFrom] (odsIn) -- (odsIn -| regFileOut);

    \draw[blockArrowTo] (flagsOut) -- (flagsOut -| regFileIn);
    \draw[blockArrowFrom] (flagsIn) -- (flagsIn -| regFileOut);

    \draw[blockArrowTo] (idsLeftOut) -- (idsLeftOut -| regFileIn);
    \draw[blockArrowFrom] (idsLeftIn) -- (idsLeftIn -| regFileOut);

    \draw[blockArrowTo] (odsOd0) -- (ctrlSId0);
    \draw[blockArrowTo] (odsOd1) -- (ctrlSId1);
    \draw[blockArrowTo] (odsOd2) -- (ctrlSId2);
    \draw[blockArrowTo] (odsOd3) -- (ctrlSId3);
    \draw[blockArrowTo] (odsOd4) -- (ctrlSId4);
    \draw[blockArrowTo] (odsOd5) -- (ctrlSId5);
    \draw[blockArrowTo] (odsOd6) -- (ctrlSId6);

    \draw[blockArrowFrom] (idsId1) -- (idsId1 |- ctrlNOd0);
    \draw[blockArrowFrom] (idsId2) -- (idsId2 |- ctrlNOd1);
    \draw[blockArrowFrom] (idsId3) -- (idsId3 |- ctrlNOd2);
    \draw[blockArrowFrom] (idsId4) -- (idsId4 |- ctrlNOd3);
    \draw[blockArrowFrom] (idsId5) -- (idsId5 |- ctrlNOd4);
    \draw[blockArrowFrom] (idsId6) -- (idsId6 |- ctrlNOd5);
    \draw[blockArrowFrom] (idsId7) -- (idsId7 |- ctrlNOd6);
}

\tikzpic{processingElement}{1}{
    \pic (box) at (0.5cm + 7.625cm, 1cm) {functionalUnit={INT\,32}};
    \pic (box) at (0.5cm + 7.625cm, 3.625cm) {functionalUnit={FP32/8$\times$4}};
    \pic (box) at (0.5cm + 7.625cm, 6.25cm) {functionalUnit={FP32/8$\times$4}};
    \pic (box) at (0.5cm + 7.625cm, 8.865cm) {functionalUnit={FP32/8$\times$4}};

    \pic (box) at (0.5cm + 0, 1.125cm) {registerFile={4}};

    \pic (box) at (0.5cm + 7.125cm+7.625cm, 1.125cm) {ctrlRegisterFile={4}};
}

\tikzpic{simpleIterationSpace}{7}{
    %

    \coordinate (a) at (0, 0);
    \coordinate (b) at (2 * #2 * #3, 0);
    \coordinate (c) at (0, 2 * #2 * #4);
    \coordinate (d) at (2 * #2 * #3, 2 * #2 * #4);

    \coordinate (e) at (#5 * #2 - #2, #5 * #2 / 2 - #2 / 2);
    \coordinate (f) at (#5 * #2 - #2 + 2 * #2 * #3, #5 * #2 / 2 - #2 / 2);
    \coordinate (g) at (#5 * #2 - #2, #5 * #2 / 2 - #2 / 2 + 2 * #2 * #4);
    \coordinate (h) at (#5 * #2 - #2 + 2 * #2 * #3, #5 * #2 / 2 - #2 / 2 + 2 * #2 * #4);

    \colorlet{borderColor}{lightgray}

    \fill[borderColor, opacity=0.75] (a) -- (e) -- (g) -- (c) -- (a);

    \fill[borderColor, opacity=0.75] (a) -- (b) -- (f) -- (e) -- (a);

    \fill[borderColor, opacity=0.75] (e) -- (f) -- (h) -- (g) -- (e);

    \draw[-] (a) -- (e);
    \draw[-] (e) -- (g);

    \foreach \k in {#5,...,1} {
        \foreach \i in {1,...,#3} {
            \foreach \j in {1,...,#4} {
                \draw[drop shadow={shadow xshift=0.015cm, shadow yshift=-0.015cm, color=black}, fill=white] ({#2 + (\i - 1) * #2 * 2 + (\k - 1) * #2}, {#2 + (\j - 1) * #2 * 2 + (\k - 1) * #2 / 2}) circle (#1);
            }
        }
    }

    \fill[borderColor, opacity=0.75] (b) -- (f) -- (h) -- (d) -- (b);

    \fill[borderColor, opacity=0.75] (c) -- (d) -- (h) -- (g) -- (c);


    \draw[-] (a) -- (b);
    \draw[-] (a) -- (c);

    \draw[-] (b) -- (f);
    \draw[-] (b) -- (d);

    \draw[-] (h) -- (d);
    \draw[-] (h) -- (g);
    \draw[-] (h) -- (f);

    \draw[-] (c) -- (g);
    \draw[-] (c) -- (d);

    \foreach \xdim\ydim\zdim in {#6} {
        \foreach \xscale\yscale\zscale in {#7} {
            \draw[blockArrowTo] (-#2, -#2) -- node[midway, below] {\xdim} +({2 * #2 + 2 * #2 * #3 * \xscale}, 0);
            \draw[blockArrowTo] (-#2, -#2) -- node[midway, left] {\ydim} +(0, {2 * #2 + 2 * #2 * #4 * \yscale});
            \draw[blockArrowTo] (-#2, 2 * #2 * #4 * \yscale+ #2) -- node[midway, above, sloped] {\zdim} +({#5 * #2 * \zscale - #2 + 2 * #2}, {#5 * #2 * \zscale / 2 - #2 / 2 + #2});
        }
    }
}

\tikzpic{tiledSimpleIterationSpace}{7}{
    \foreach \xscale\yscale\zscale in {#7} {
        \foreach \z in {\zscale,...,\zscale} {
            \foreach \x in {1,...,\xscale} {
                \foreach \y in {1,...,\yscale} {
                    \pic (space) at ({2 * #2 * #3 * (\x - 1) + (#5 - 1) * #2 * (\z + 1) }, {2 * #2 * #4 * (\y - 1) + (#5 - 1) * #2 * (\z + 1) / 2}) {blockPic={PE}{2 * #2 * #3}{2 * #2 * #4}{///}{///}};

                    \draw[blockArrowFrom] ({2 * #2 * #3 * (\x - 1) + (#5 - 1) * #2 * (\z + 1) }, {2 * #2 * #4 * (\y - 1) + (#5 - 1) * #2 * (\z + 1) / 2}) -- +(-{(#5 - 1) * #2}, -{(#5 - 1) * #2 / 2});
                    \draw[blockArrowFrom] ({2 * #2 * #3 * (\x - 1) + (#5 - 1) * #2 * (\z + 1) + 2 * #2 * #3}, {2 * #2 * #4 * (\y - 1) + (#5 - 1) * #2 * (\z + 1) / 2}) -- +(-{(#5 - 1) * #2}, -{(#5 - 1) * #2 / 2});
                    \draw[blockArrowFrom] ({2 * #2 * #3 * (\x - 1) + (#5 - 1) * #2 * (\z + 1) + 2 * #2 * #3}, {2 * #2 * #4 * (\y - 1) + (#5 - 1) * #2 * (\z + 1) / 2 + 2 * #2 * #4}) -- +(-{(#5 - 1) * #2}, -{(#5 - 1) * #2 / 2});
                    \draw[blockArrowFrom] ({2 * #2 * #3 * (\x - 1) + (#5 - 1) * #2 * (\z + 1)}, {2 * #2 * #4 * (\y - 1) + (#5 - 1) * #2 * (\z + 1) / 2 + 2 * #2 * #4}) -- +(-{(#5 - 1) * #2}, -{(#5 - 1) * #2 / 2});

                }
            }
        }

        \foreach \z in {\zscale,...,1} {
            \foreach \x in {1,...,\xscale} {
                \foreach \y in {1,...,\yscale} {
                    \ifnum\x=1
                        \ifnum\y=1
                            \ifnum\z=1
                                \pic (space) at ({2 * #2 * #3 * (\x - 1) + (#5 - 1) * #2 * (\z - 1) }, {2 * #2 * #4 * (\y - 1) + (#5 - 1) * #2 * (\z - 1) / 2}) {simpleIterationSpace={#1}{#2}{#3}{#4}{#5}{#6}{#7}};
                            \else
                                \pic (space) at ({2 * #2 * #3 * (\x - 1) + (#5 - 1) * #2 * (\z - 1) }, {2 * #2 * #4 * (\y - 1) + (#5 - 1) * #2 * (\z - 1) / 2}) {simpleIterationSpace={#1}{#2}{#3}{#4}{#5}{}{}};
                            \fi
                        \else
                            \pic (space) at ({2 * #2 * #3 * (\x - 1) + (#5 - 1) * #2 * (\z - 1) }, {2 * #2 * #4 * (\y - 1) + (#5 - 1) * #2 * (\z - 1) / 2}) {simpleIterationSpace={#1}{#2}{#3}{#4}{#5}{}{}};
                        \fi
                    \else
                        \pic (space) at ({2 * #2 * #3 * (\x - 1) + (#5 - 1) * #2 * (\z - 1) }, {2 * #2 * #4 * (\y - 1) + (#5 - 1) * #2 * (\z - 1) / 2}) {simpleIterationSpace={#1}{#2}{#3}{#4}{#5}{}{}};
                    \fi
                }
            }
        }
    }

    \node[inner sep=0] at ({2 * #2 * #3 * (8) + #5 * #2 * (1) - #2}, {2 * #2 * #4 * (8) + #5 * #2 * (1) / 2 - #2 / 2}) (NorthEast) {};
    \node[inner sep=0] at ({2 * #2 * #3 * (0) + #5 * #2 * (1) - #2}, {2 * #2 * #4 * (8) + #5 * #2 * (1) / 2 - #2 / 2}) (NorthWest) {};
    \node[inner sep=0] at ({2 * #2 * #3 * (8) + #5 * #2 * (1) - #2}, {2 * #2 * #4 * (0) + #5 * #2 * (1) / 2 - #2 / 2}) (SouthEast) {};
    \node[inner sep=0] at ({2 * #2 * #3 * (0) + #5 * #2 * (1) - #2}, {2 * #2 * #4 * (0) + #5 * #2 * (1) / 2 - #2 / 2}) (SouthWest) {};
}

\tikzpic{nodeGrid}{8}{
    \pgfmathsetlengthmacro{\cellSize}{#1};
    \pgfmathsetlengthmacro{\nodeSize}{#2};
    \pgfmathsetlengthmacro{\textSize}{#3};
    \pgfmathsetlengthmacro{\arrowLineWidth}{#5};

    \foreach \row [count=\j] in {#4} {
        \foreach \cell [count=\i] in \row {
            \foreach \text\label\foreground\background in \cell {
                \colorlet{leftColor}{\background}
                \colorlet{rightColor}{leftColor!80}

                \node[text=\foreground, font=\fontsize{\textSize}{8}\selectfont, drop shadow={shadow xshift=0.015cm, shadow yshift=-0.015cm, color=black}, draw, minimum width=\nodeSize, inner sep=0, shading=axis, left color=leftColor, right color=rightColor, shading angle=0, circle] (\label) at (-\cellSize / 2 + \i * \cellSize, \cellSize / 2 - \j * \cellSize) {};
                \node[text=\foreground, font=\fontsize{\textSize}{8}\selectfont] at (-\cellSize / 2 + \i * \cellSize, \cellSize / 2 - \j * \cellSize) {\textbf{\text}};
            }
        }
    }

    \foreach \arrow in {#6} {
        \foreach \start\end\colorName\labelText\outAngle\inAngle\loose\styleThings in \arrow {
            \colorlet{drawColor}{\colorName}

            \ifthenelse{\equal{\outAngle}{}}{
                \draw[-{Latex[length=4*\arrowLineWidth]}, line width=\arrowLineWidth, draw=drawColor, \styleThings] (\start) to (\end);
                \ifthenelse{\equal{\labelText}{}}{
                    \draw[-{Latex[length=4*\arrowLineWidth]}, line width=\arrowLineWidth, draw=drawColor, \styleThings] (\start) to (\end);
                }{
                    \draw[-{Latex[length=4*\arrowLineWidth]}, line width=\arrowLineWidth, draw=drawColor, \styleThings] (\start) to node[pos=0.5, fill=white, text=drawColor, rounded corners=0.2cm] {\labelText} (\end);
                }
            }{
                \ifthenelse{\equal{\labelText}{}}{
                    \draw[-{Latex[length=4*\arrowLineWidth]}, looseness=\loose, line width=\arrowLineWidth, draw=drawColor, \styleThings] (\start) to[out=\outAngle, in=\inAngle] (\end);
                }{
                    \draw[-{Latex[length=4*\arrowLineWidth]}, looseness=\loose, line width=\arrowLineWidth, draw=drawColor, \styleThings] (\start) to[out=\outAngle, in=\inAngle] node[pos=0.5, fill=white, text=drawColor, rounded corners=0.2cm] {\labelText} (\end);
                }
            }
        }
    }

    \foreach \border in {#7} {
        \foreach \startX\startY\endX\endY\styleThings in \border {
            \draw[\styleThings] (\cellSize * \startX, -\cellSize * \startY) -- (\cellSize * \endX, -\cellSize * \endY);
        }
    }
    \foreach \label in {#8} {
        \foreach \startX\startY\endX\endY\labelText\labelStyle in \label {
            \node[\labelStyle, align=center] at ({\cellSize * (\startX + (\endX-\startX) * 0.5)}, {\cellSize * -(\startY + (\endY-\startY) * 0.5)}) {\labelText};
        }
    }
}

\tikzpic{dfgPic}{1}{
    \definecolor{darkPurple}{RGB}{96, 25, 134}   
    \definecolor{darkBlue}{RGB}{0, 59, 111}      
    \definecolor{darkBrown}{RGB}{102, 51, 0}     
    \definecolor{darkGray}{RGB}{71, 79, 82}   
    \definecolor{darkRed}{RGB}{177, 0, 18}
    \definecolor{darkGreen}{RGB}{0, 147, 146}
    \definecolor{darkOrange}{RGB}{255, 111, 0}

    \pic at (0cm, 0cm) {nodeGrid={1.25cm}{0.75cm}{
        {{+/Addr0/white/darkBlue},{},{+/Addr1/white/darkBlue}},
        {{L/Load0/white/darkBlue},{},{L/Load1/white/darkBlue}},
        {{},{\texttimes/Mul/white/darkRed}, {+/ItAdd/white/darkPurple}},
        {{},{+/Sum/white/darkRed}, {=/Cmp/white/darkPurple}},
        {{},{S/Store/white/darkGreen},{}}%
    }{0.03cm}{
        {Addr0/Addr0/darkBlue/45/0/7},
        {Addr1/Addr1/darkBlue/45/0/7},
        {Sum/Sum/darkRed/45/0/7},
        {ItAdd/ItAdd/darkPurple/45/0/7},
        {Addr0/Load0/darkBlue},
        {Addr1/Load1/darkBlue},
        {Load0/Mul/darkBlue},
        {Load1/Mul/darkBlue},
        {Mul/Sum/darkRed},
        {Sum/Store/darkRed},
        {ItAdd/Cmp/darkPurple},
        {Cmp/Store/darkPurple}%
    }};
}

\tikzpic{cgraMappingPic}{1}{
    \definecolor{darkPurple}{RGB}{96, 25, 134}   
    \definecolor{darkBlue}{RGB}{0, 59, 111}      
    \definecolor{darkBrown}{RGB}{102, 51, 0}     
    \definecolor{darkGray}{RGB}{71, 79, 82}   
    \definecolor{darkRed}{RGB}{177, 0, 18}
    \definecolor{darkGreen}{RGB}{0, 147, 146}
    \definecolor{darkOrange}{RGB}{255, 111, 0}
    
    \foreach \j in {0,...,2} {
        \foreach \i in {0,...,2} {
            \pic (a) at (2cm * \i + 1cm - 1.8cm/2, -2cm * \j - 1cm - 1.8cm/2) {blockPic={}{1.8cm}{1.8cm}{///}{////}};
            \node[anchor=south east] at (2cm * \i + 1cm - 1.8cm/2+1.8cm, -2cm * \j - 1cm - 1.8cm/2) {PE};
        }
    }

    \pgfmathsetlengthmacro{\cgraLeftTopX}{1cm - 1.8cm/2};
    \pgfmathsetlengthmacro{\cgraLeftTopY}{-1cm + 1.8cm/2};

    \pgfmathsetlengthmacro{\cgraRightBotX}{\cgraLeftTopX + 6cm - 0.2cm};
    \pgfmathsetlengthmacro{\cgraRightBotY}{\cgraLeftTopY - 6cm + 0.2cm};

    \pic (AGE) at (\cgraRightBotX + 0.2cm, \cgraRightBotY) {rotatedBlockPic={Memory}{0.9cm}{2cm*3-0.2cm}{///6}{///{p0, p2, p3, p4, p5, p6}}};
    \pic (AGW) at (\cgraLeftTopX - 0.2cm - 0.9cm, \cgraRightBotY) {rotatedBlockPic={Memory}{0.9cm}{2cm*3-0.2cm}{/6//}{{}/{p0, p2, p3, p4, p5, p6}//}};

    \pic (cgra) at (0, 0) {nodeGrid={2.0cm}{0.75cm}{
        {{L/Load0/white/darkBlue}, {+/Addr0/white/darkBlue}, {+/ItAdd/white/darkPurple}},
        {{L/Load1/white/darkBlue}, {+/Addr1/white/darkBlue}, {=/Cmp/white/darkPurple}},
        {{\texttimes/Mul/white/darkRed},   {+/Sum/white/darkRed},    {S/Store/white/darkGreen}}%
    }{0.02cm}{
        {Addr0/Addr0/darkBlue/90/45/5},
        {Addr1/Addr1/darkBlue/90/45/5},
        {Sum/Sum/darkRed/90/45/5},
        {ItAdd/ItAdd/darkPurple/90/45/5},
        {Addr0/Load0/darkBlue},
        {Addr1/Load1/darkBlue},
        {Load0/Mul/darkBlue/245/115/1},
        {Load1/Mul/darkBlue},
        {Mul/Sum/darkRed},
        {Sum/Store/darkRed},
        {ItAdd/Cmp/darkPurple},
        {Cmp/Store/darkPurple}%
    }};
    \draw[{Latex[length=4*0.02cm]}-{Latex[length=4*0.02cm]}, line width=0.02cm, draw=darkBlue] (cgraLoad0) -- (cgraLoad0 -| AGWp6);
    \draw[{Latex[length=4*0.02cm]}-{Latex[length=4*0.02cm]}, line width=0.02cm, draw=darkBlue] (cgraLoad1) -- (cgraLoad1 -| AGWp4);
    \draw[-{Latex[length=4*0.02cm]}, line width=0.02cm, draw=darkGreen] (cgraStore) -- (cgraStore -| AGEp0);
}

\tikzpic{tcpaMappingPic}{1}{
    \definecolor{darkPurple}{RGB}{96, 25, 134}   
    \definecolor{darkBlue}{RGB}{0, 59, 111}      
    \definecolor{darkBrown}{RGB}{102, 51, 0}     
    \definecolor{darkGray}{RGB}{71, 79, 82}   
    \definecolor{darkRed}{RGB}{177, 0, 18}
    \definecolor{darkGreen}{RGB}{0, 147, 146}
    \definecolor{darkOrange}{RGB}{255, 111, 0}
    \foreach \j in {0,...,2} {
        \foreach \i in {0,...,2} {
            \pic (a) at (2cm * \i + 1cm - 1.8cm/2, -2cm * \j - 1cm - 1.8cm/2) {blockPic={}{1.8cm}{1.8cm}{///}{////}};
        }
    }
    \pgfmathsetlengthmacro{\tcpaLeftTopX}{1cm - 1.8cm/2};
    \pgfmathsetlengthmacro{\tcpaLeftTopY}{-1cm + 1.8cm/2};

    \pgfmathsetlengthmacro{\tcpaRightBotX}{\tcpaLeftTopX + 6cm - 0.2cm};
    \pgfmathsetlengthmacro{\tcpaRightBotY}{\tcpaLeftTopY - 6cm + 0.2cm};

    \pic (GC) at (\tcpaLeftTopX - 0.2cm - 0.9cm, \tcpaLeftTopY + 0.2cm) {blockPic={GC}{0.9cm}{0.9cm}{///}{{}/{}//}};

    \pic (AGN) at (\tcpaLeftTopX, \tcpaLeftTopY + 0.2cm) {blockPic={I/O Buffer\,\&\,Address Generators}{2cm*3-0.2cm}{0.9cm}{///}{{}/{}/{}/}};
    \pic (AGS) at (\tcpaLeftTopX, \tcpaRightBotY - 1.1cm) {blockPic={I/O Buffer\,\&\,Address Generators}{2cm*3-0.2cm}{0.9cm}{///}{{}/{}//}};

    \pic (AGE) at (\tcpaRightBotX + 0.2cm, \tcpaRightBotY) {rotatedBlockPic={I/O Buffer\,\&\,Address Generators}{0.9cm}{2cm*3-0.2cm}{///}{///{}}};
    \pic (AGW) at (\tcpaLeftTopX - 0.2cm - 0.9cm, \tcpaRightBotY) {rotatedBlockPic={I/O Buffer\,\&\,Address Generators}{0.9cm}{2cm*3-0.2cm}{///}{{}/{}//}};

    \foreach \v in {0,...,2} {
        \pic (AGN) at (\tcpaLeftTopX, \tcpaLeftTopY + 0.2cm - \v * 2cm) {emptyBlockPic={I/O Buffer\&Address Generators}{2cm*3-0.2cm}{0.9cm}{//24/}{{}/{}/{p0, p1, p2, p3, p4, p5, p6, p7, p8, p9, p10, p11, p12, p13, p14, p15, p16, p17, p18, p19, p20, p21, p22, p23}/}};
        \pic (AGE) at (\tcpaRightBotX + 0.2cm, \tcpaRightBotY - \v * 2cm) {emptyBlockPic={I/O Buffer\&Address Generators}{0.9cm}{2cm*3-0.2cm}{///6}{///{p0, p2, p3, p4, p5, p6}}};

        \pic (tcpa) at (0cm, -0.25cm - \v * 2cm) {nodeGrid={2.0cm/4}{0.1cm}{
            {{\normalsize{\texttimes}/Mul0/white/darkRed},
            {\normalsize{\texttimes}/Mul1/white/darkRed},
            {\normalsize{\texttimes}/Mul2/white/darkRed},
            {\normalsize{\texttimes}/Mul3/white/darkRed},
            {\normalsize{\texttimes}/Mul4/white/darkRed},
            {\normalsize{\texttimes}/Mul5/white/darkRed},
            {\normalsize{\texttimes}/Mul6/white/darkRed},
            {\normalsize{\texttimes}/Mul7/white/darkRed},
            {\normalsize{\texttimes}/Mul8/white/darkRed},
            {\normalsize{\texttimes}/Mul9/white/darkRed},
            {\normalsize{\texttimes}/Mul10/white/darkRed},
            {\normalsize{\texttimes}/Mul11/white/darkRed}},
            {},
            {{\normalsize{+}/Add0/white/darkRed},
            {\normalsize{+}/Add1/white/darkRed},
            {\normalsize{+}/Add2/white/darkRed},
            {\normalsize{+}/Add3/white/darkRed},
            {\normalsize{+}/Add4/white/darkRed},
            {\normalsize{+}/Add5/white/darkRed},
            {\normalsize{+}/Add6/white/darkRed},
            {\normalsize{+}/Add7/white/darkRed},
            {\normalsize{+}/Add8/white/darkRed},
            {\normalsize{+}/Add9/white/darkRed},
            {\normalsize{+}/Add10/white/darkRed},
            {\normalsize{+}/Add11/white/darkRed}}%
        }{0.02cm}{
            {Mul0/Add0/darkRed},
            {Mul1/Add1/darkRed},
            {Mul2/Add2/darkRed},
            {Mul3/Add3/darkRed},
            {Mul4/Add4/darkRed},
            {Mul5/Add5/darkRed},
            {Mul6/Add6/darkRed},
            {Mul7/Add7/darkRed},
            {Mul8/Add8/darkRed},
            {Mul9/Add9/darkRed},
            {Mul10/Add10/darkRed},
            {Mul11/Add11/darkRed},
            {Add0/Add1/darkRed},
            {Add1/Add2/darkRed},
            {Add2/Add3/darkRed},
            {Add3/Add4/darkRed},
            {Add4/Add5/darkRed},
            {Add5/Add6/darkRed},
            {Add6/Add7/darkRed},
            {Add7/Add8/darkRed},
            {Add8/Add9/darkRed},
            {Add9/Add10/darkRed},
            {Add10/Add11/darkRed}%
        }};

        \foreach \i in {0,...,11} {
            \draw[-{Latex[length=4*0.02cm]}, line width=0.02cm, draw=darkPurple] (\tcpaLeftTopX - 0.2cm, \tcpaLeftTopY + 0.2cm) -- +(0.1cm, -0.1cm) -- +(0.1cm, -2.1cm - \v * 2cm) -| (tcpaAdd\i);
        }

        \draw[-{Latex[length=4*0.02cm]}, line width=0.02cm, draw=darkBlue] (AGNp0) to[out=270, in=110] (tcpaMul0);
        \draw[-{Latex[length=4*0.02cm]}, line width=0.02cm, draw=darkBlue] (AGNp1) to[out=270, in=70]  (tcpaMul0);
        \draw[-{Latex[length=4*0.02cm]}, line width=0.02cm, draw=darkBlue] (AGNp2) to[out=270, in=110] (tcpaMul1);
        \draw[-{Latex[length=4*0.02cm]}, line width=0.02cm, draw=darkBlue] (AGNp3) to[out=270, in=70]  (tcpaMul1);
        \draw[-{Latex[length=4*0.02cm]}, line width=0.02cm, draw=darkBlue] (AGNp4) to[out=270, in=110] (tcpaMul2);
        \draw[-{Latex[length=4*0.02cm]}, line width=0.02cm, draw=darkBlue] (AGNp5) to[out=270, in=70]  (tcpaMul2);
        \draw[-{Latex[length=4*0.02cm]}, line width=0.02cm, draw=darkBlue] (AGNp6) to[out=270, in=110] (tcpaMul3);
        \draw[-{Latex[length=4*0.02cm]}, line width=0.02cm, draw=darkBlue] (AGNp7) to[out=270, in=70]  (tcpaMul3);
        \draw[-{Latex[length=4*0.02cm]}, line width=0.02cm, draw=darkBlue] (AGNp8) to[out=270, in=110] (tcpaMul4);
        \draw[-{Latex[length=4*0.02cm]}, line width=0.02cm, draw=darkBlue] (AGNp9) to[out=270, in=70]  (tcpaMul4);
        \draw[-{Latex[length=4*0.02cm]}, line width=0.02cm, draw=darkBlue] (AGNp10) to[out=270, in=110] (tcpaMul5);
        \draw[-{Latex[length=4*0.02cm]}, line width=0.02cm, draw=darkBlue] (AGNp11) to[out=270, in=70]  (tcpaMul5);
        \draw[-{Latex[length=4*0.02cm]}, line width=0.02cm, draw=darkBlue] (AGNp12) to[out=270, in=110] (tcpaMul6);
        \draw[-{Latex[length=4*0.02cm]}, line width=0.02cm, draw=darkBlue] (AGNp13) to[out=270, in=70]  (tcpaMul6);
        \draw[-{Latex[length=4*0.02cm]}, line width=0.02cm, draw=darkBlue] (AGNp14) to[out=270, in=110] (tcpaMul7);
        \draw[-{Latex[length=4*0.02cm]}, line width=0.02cm, draw=darkBlue] (AGNp15) to[out=270, in=70]  (tcpaMul7);
        \draw[-{Latex[length=4*0.02cm]}, line width=0.02cm, draw=darkBlue] (AGNp16) to[out=270, in=110] (tcpaMul8);
        \draw[-{Latex[length=4*0.02cm]}, line width=0.02cm, draw=darkBlue] (AGNp17) to[out=270, in=70]  (tcpaMul8);
        \draw[-{Latex[length=4*0.02cm]}, line width=0.02cm, draw=darkBlue] (AGNp18) to[out=270, in=110] (tcpaMul9);
        \draw[-{Latex[length=4*0.02cm]}, line width=0.02cm, draw=darkBlue] (AGNp19) to[out=270, in=70]  (tcpaMul9);
        \draw[-{Latex[length=4*0.02cm]}, line width=0.02cm, draw=darkBlue] (AGNp20) to[out=270, in=110] (tcpaMul10);
        \draw[-{Latex[length=4*0.02cm]}, line width=0.02cm, draw=darkBlue] (AGNp21) to[out=270, in=70]  (tcpaMul10);
        \draw[-{Latex[length=4*0.02cm]}, line width=0.02cm, draw=darkBlue] (AGNp22) to[out=270, in=110] (tcpaMul11);
        \draw[-{Latex[length=4*0.02cm]}, line width=0.02cm, draw=darkBlue] (AGNp23) to[out=270, in=70]  (tcpaMul11);

        \draw[-{Latex[length=4*0.02cm]}, line width=0.02cm, draw=darkRed] (tcpaAdd11) to[out=45, in=225] (AGEp6);
    }

}
\tikzpic{mappingComparison}{1}{
    \pic (dfg) at (0, 0) {dfgPic={}};
    \pic (cgra) at (6, 0) {cgraMappingPic={}};
    \pic (tcpa) at (15, 0) {tcpaMappingPic={}};
}
\tikzpic{verticalRangeNode}{6}{
    \pgfmathsetlengthmacro{\width}{#1};
    \pgfmathsetlengthmacro{\height}{#2};
    \pgfmathsetlengthmacro{\curveWidth}{#3};
    \pgfmathsetlengthmacro{\curveStart}{\width - \curveWidth};
    \pgfmathsetlengthmacro{\curveSlot}{\height/8};

    \colorlet{textColor}{#5}
    \colorlet{leftColor}{#4}
    \colorlet{rightColor}{leftColor}

    \draw [fill, drop shadow={shadow xshift=0.02cm, shadow yshift=-0.02cm, color=black}, shading=axis, left color=leftColor, right color=rightColor, shading angle=0, smooth, rounded corners=2 * \curveSlot * 0.10] (0, 0) -- (\curveStart + 0, 0) -- (\curveStart + \curveWidth/2, 0.5 * \curveSlot) -- (\curveStart + 0.75*\curveWidth, 1 * \curveSlot) -- (\curveStart + \curveWidth/2, 2 * \curveSlot) -- (\curveStart + 0.25*\curveWidth, 3 * \curveSlot) -- (\curveStart + \curveWidth/2, 2 * \curveSlot * 1.8) {[sharp corners] -- (\curveStart + \curveWidth, 4 * \curveSlot)} -- (\curveStart + \curveWidth/2, 2 * \curveSlot * 2.2) -- (\curveStart + 0.25*\curveWidth, 5 * \curveSlot) -- (\curveStart + \curveWidth/2, 6 * \curveSlot) -- (\curveStart + 0.75*\curveWidth, 7 * \curveSlot) -- (\curveStart + \curveWidth/2, 7.5 * \curveSlot) -- (\curveStart + 0, 8 * \curveSlot) -- (0, \height) -- cycle;

    \node[anchor=west, align=center, text=textColor] at (\width, 0.5 * \height) {#6};
}

\tikzpic{verticalRangeNodeWithExtension}{8}{
    \pgfmathsetlengthmacro{\width}{#1};
    \pgfmathsetlengthmacro{\height}{#2};
    \pgfmathsetlengthmacro{\curveWidth}{#3};
    \pgfmathsetlengthmacro{\curveStart}{\width - \curveWidth};
    \pgfmathsetlengthmacro{\curveSlot}{\height/8};
    \pgfmathsetlengthmacro{\extensionWidth}{#7};
    \pgfmathsetlengthmacro{\extensionHeight}{#8};

    \colorlet{textColor}{#5}
    \colorlet{leftColor}{#4}
    \colorlet{rightColor}{leftColor}

    \draw [fill, drop shadow={shadow xshift=0.02cm, shadow yshift=-0.02cm, color=black}, shading=axis, left color=leftColor, right color=rightColor, shading angle=0, smooth, rounded corners=2 * \curveSlot * 0.10] (0, -\extensionHeight) -- (\extensionWidth, -\extensionHeight) -- (\extensionWidth, 0) -- (\curveStart + 0, 0) -- (\curveStart + \curveWidth/2, 0.5 * \curveSlot) -- (\curveStart + 0.75*\curveWidth, 1 * \curveSlot) -- (\curveStart + \curveWidth/2, 2 * \curveSlot) -- (\curveStart + 0.25*\curveWidth, 3 * \curveSlot) -- (\curveStart + \curveWidth/2, 2 * \curveSlot * 1.8) {[sharp corners] -- (\curveStart + \curveWidth, 4 * \curveSlot)} -- (\curveStart + \curveWidth/2, 2 * \curveSlot * 2.2) -- (\curveStart + 0.25*\curveWidth, 5 * \curveSlot) -- (\curveStart + \curveWidth/2, 6 * \curveSlot) -- (\curveStart + 0.75*\curveWidth, 7 * \curveSlot) -- (\curveStart + \curveWidth/2, 7.5 * \curveSlot) -- (\curveStart + 0, 8 * \curveSlot) -- (0, \height) -- cycle;

    \node[anchor=west, align=center, text=textColor] at (\width, 0.5 * \height) {#6};
}

\tikzpic{textRangeBlock}{7}{
    \pgfmathsetlengthmacro{\width}{#1};
    \pgfmathsetlengthmacro{\height}{#2};
    \pgfmathsetlengthmacro{\numRows}{#3};
    \pgfmathsetlengthmacro{\textWidth}{#4};
    \pgfmathsetlengthmacro{\rowHeight}{\height / \numRows};

    \foreach \foreground\background\label in {#5} {
        \colorlet{textColor}{\foreground}
        \colorlet{leftColor}{\background}
        \colorlet{rightColor}{leftColor!80}

        \pic at (0, 0) {verticalRangeNode={\width}{\height}{0.5cm}{\background}{\label}{#7}};
        \foreach \line [count=\i] in {#6} {
            \node[anchor=south west, align=center, minimum height=\rowHeight, text=textColor, inner sep=0] at (0, {\rowHeight * (\numRows - \i)}) {{\line}};
        }
    }
}
\tikzpic{textRangeBlockWithExtension}{9}{
    \pgfmathsetlengthmacro{\width}{#1};
    \pgfmathsetlengthmacro{\height}{#2};
    \pgfmathsetlengthmacro{\numRows}{#3};
    \pgfmathsetlengthmacro{\textWidth}{#4};
    \pgfmathsetlengthmacro{\rowHeight}{\height / \numRows};

    \foreach \foreground\background\label in {#5} {
        \colorlet{textColor}{\foreground}
        \colorlet{leftColor}{\background}
        \colorlet{rightColor}{leftColor!80}

        \pic at (0, 0) {verticalRangeNodeWithExtension={\width}{\height}{0.5cm}{\background}{\label}{#7}{#8}{#9}};
        \foreach \line [count=\i] in {#6} {
            \node[anchor=south west, align=center, minimum height=\rowHeight, text=textColor, inner sep=0] at (0, {\rowHeight * (\numRows - \i)}) {\textbf{\line}};
        }
    }
}
\tikzpic{loopDescription}{1}{
    \definecolor{darkPurple}{RGB}{96, 25, 134}   
    \definecolor{darkBlue}{RGB}{0, 59, 111}      
    \definecolor{darkBrown}{RGB}{102, 51, 0}     
    \definecolor{darkGray}{RGB}{71, 79, 82}   
    \definecolor{darkRed}{RGB}{177, 0, 18}
    \definecolor{darkGreen}{RGB}{0, 147, 146}
    \definecolor{darkOrange}{RGB}{255, 111, 0}

    \definecolor{normalGray}{HTML}{f4f4f4};

    \pic at (0, 1.25*2.5) {textRangeBlockWithExtension={10.5cm}{2.5cm}{4}{3cm}{black/normalGray/black}{
            {\hspace*{0.25cm}for $0 \leq {i}_{0} < N_0:$},
            {\hspace*{0.75cm}for $0 \leq {i}_{1} < N_1:$},
            {\hspace*{1.25cm}$\dots$},
            {\hspace*{1.25cm}for $0 \leq {i}_{7} < N_7:$}%
    }{8 Dimensional\\ Iteration Space}{1.5cm}{1.25cm/2+2.5cm}};

    \pic at (1.5cm, 1.25*2.0) {textRangeBlock={9.0cm}{1.25cm/2}{1}{3cm}{black/normalGray/black}{
            {\hspace*{0.25cm}{x[${i}_{0}, \dots, {i}_{7}$] = X[$a_0{i}_{0} + \dots + a_7{i}_{7}$]}}%
    }{Input Dep.}};

    \pic at (1.5cm, 1.25*1.5) {textRangeBlock={9.0cm}{1.25cm/2}{1}{3cm}{black/normalGray/black}{
            {\hspace*{0.25cm}{Z[$a_0{i}_{0} + \dots + a_7{i}_{7}$] = z[${i}_{0}, \dots, {i}_{7}$]}}%
    }{Output Dep.}};

    \pic at (1.5cm, 1.25) {textRangeBlock={9.0cm}{1.25cm/2}{1}{3cm}{black/normalGray/black}{
            {\hspace*{0.25cm}{u[${i}_{0}, \dots, {i}_{7}$] = x[${i}_{0}, \dots, {i}_{7}$]\ $\times$\ y[${i}_{0}, \dots, {i}_{7}$]}}%
    }{Intra-iteration Dep.}};
    \pic at (1.5cm, 1.25/2) {textRangeBlock={9.0cm}{1.25cm/2}{1}{3cm}{black/normalGray/black}{
            {\hspace*{0.25cm}{s[${i}_{0}, \dots, {i}_{7}$] = s[${i}_{0} - d_0, \dots, {i}_{7} - d_7$]}}%
    }{Loop-carried Dep.}};

    \pic at (1.5cm, 0) {textRangeBlock={9.0cm}{1.25cm/2}{1}{3cm}{black/normalGray/black}{
            {\hspace*{0.25cm}{z[${i}_{0}, \dots, {i}_{7}$] = s[${i}_{0}, \dots, {i}_{7}$] if $i_7 = N_7 - 1$}}%
    }{Conditionals}};
}

\tikzpic{computeCapability}{1}{
    \pic (itSpace) at (14.5cm, 2cm) {tiledSimpleIterationSpace={0.13cm}{0.15cm}{2}{2}{8}{{$i_0$}/{$i_1$}/{$i_2, \dots, i_7$}}{8/8/1}};
    \coordinate (selectedIt) at (1.0cm + 0.2cm * 8 * 2 * 2 - 0.2cm, 1.0cm + 0.2cm * 8 * 2 * 2 - 0.2cm);

    \pic  at (0.0cm, 2cm) {loopDescription={}};
}

\tikzpic{transBox}{3} {
    \node[anchor=south west, rounded corners=0.05cm, minimum width=#2, minimum height=#3, inner sep=0, draw, thick, white, opacity=0.75] at (0, 0) {#1};
}
\tikzpic{transBoxVert}{3} {
    \node[anchor=north west, rotate=90, rounded corners=0.05cm, minimum width=#3, minimum height=#2, inner sep=0, draw, thick, white, opacity=0.75] at (0, 0) {#1};
}
\tikzpic{dieShot}{1}{
    \node[anchor=south west] at (0, 0) { \includegraphics[width=11.5cm]{alpaca_die_cropped.jpg} };
    
    \draw[blockArrowBi] (0.53cm, 0.12cm) -- node[midway, fill=white] {\qty{3.2}{\milli\metre}} (11.25cm, 0.12cm);
    \draw[blockArrowBi] (0.15cm, 0.5cm) -- node[midway, fill=white, sloped] {\qty{3.125}{\milli\metre}} (0.15cm, 10.9cm);

    \foreach \i in {0,...,7} {
        \foreach \j in {0,...,7} {
            \pic (a) at (2.9cm+0.82cm*\i, 2.2cm+0.82cm*\j) {transBox={PE}{0.75cm}{0.7cm}};
        }
    }

    \pic (a) at (2.9, 0.8) {transBox={I/O Buffers and AGs}{8*0.82cm-0.07cm}{1.3cm}};
    \pic (a) at (2.9, 0.8+7.9) {transBox={I/O Buffers and AGs}{8*0.82cm-0.07cm}{1.2cm}};

    \pic (a) at (2.9-1.3, 2.2) {transBoxVert={I/O Buffers and AGs}{1.2cm}{8*0.82cm-0.1cm}};
    \pic (a) at (3.0+0.82*8, 2.2) {transBoxVert={I/O Buffers and AGs}{1.2cm}{8*0.82cm-0.1cm}};

    \pic (a) at (1.2, 9.4) {transBox={LION}{1.6cm}{1.1cm}};
    \pic (a) at (1.6, 8.75) {transBox={DMA}{1.2cm}{0.6cm}};
    \pic (a) at (2.9, 10) {transBox={GC}{1.2cm}{0.5cm}};

}
\tikzpic{layout}{1}{

    \node[anchor=south west] at (7pt, 7pt) { \includegraphics[width=11.5cm]{alpaca_floorplan.jpg} };
    
    \draw[opacity=0, white, blockArrowBi] (0.53cm, 0.12cm) -- (11.25cm, 0.12cm);
    \draw[opacity=0, white, blockArrowBi] (0.15cm, 0.5cm) -- (0.15cm, 10.9cm);
}

\tikzpic{arrowTableColumns}{7}{
    \pgfmathsetlengthmacro{\leftWidth}{#1};
    \pgfmathsetlengthmacro{\centerWidth}{#2};
    \pgfmathsetlengthmacro{\rightWidth}{#3};
    \pgfmathsetlengthmacro{\height}{#4};
    \pgfmathsetlengthmacro{\lineWidth}{#5};

    \pgfmathsetlengthmacro{\arrowHeight}{\lineWidth * 2.00};
    \pgfmathsetlengthmacro{\arrowWidth}{\arrowHeight * 0.5 * 3};
    \pgfmathsetlengthmacro{\arrowLineOffset}{\arrowWidth / 3 + ((\arrowWidth / 3 * 2) / (\arrowHeight / 2)) * (\lineWidth / 2)}

    \pgfmathsetlengthmacro{\boxHeight}{\height-0.5 * \arrowHeight};

    \definecolor{leftColor}{HTML}{d9d9d9};
    \definecolor{rightColor}{HTML}{d9d9d9};

    \foreach \leftName\centerName\rightName in {#6} {
        \foreach \leftColor\centerColor\rightColor in {#7} {
            \foreach \leftForeground\leftBackground in \leftColor {
                \foreach \centerForeground\centerBackground in \centerColor {
                    \foreach \rightForeground\rightBackground in \rightColor {
                        \shade[left color=\leftBackground!10, right color=\centerBackground!10, shading angle=90] (0, 0) -- +(\leftWidth, 0) -- +(\leftWidth, \boxHeight) -- +(0, \boxHeight) -- (0, 0);
                        \shade[shading=axis, left color=\centerBackground!10, right color=\centerBackground!10, shading angle=90] (\leftWidth, 0) -- +(\centerWidth, 0) -- +(\centerWidth, \boxHeight) -- +(0, \boxHeight) -- +(0, 0);
                        \shade[shading=axis, left color=\centerBackground!10, right color=\rightBackground!10, shading angle=90] (\leftWidth + \centerWidth, 0) -- +(\rightWidth, 0) -- +(\rightWidth, \boxHeight) -- +(0, \boxHeight) -- +(0, 0);

                        \draw[drop shadow={shadow xshift=0.02cm, shadow yshift=-0.02cm, color=black}] 
                            (0, \boxHeight) -- 
                            +(\arrowWidth, \arrowHeight/2) -- 
                            +(\arrowLineOffset, \lineWidth / 2) -- 
                            +(\leftWidth + \centerWidth + \rightWidth - \arrowLineOffset, \lineWidth / 2) -- 
                            +(\leftWidth + \centerWidth + \rightWidth - \arrowWidth, \arrowHeight / 2) -- 
                            +(\leftWidth + \centerWidth + \rightWidth, 0) -- 
                            +(\leftWidth + \centerWidth + \rightWidth - \arrowWidth, -\arrowHeight / 2) -- 
                            +(\leftWidth + \centerWidth + \rightWidth - \arrowLineOffset, -\lineWidth / 2) -- 
                            +(\arrowLineOffset, -\lineWidth / 2) -- 
                            +(\arrowWidth, -\arrowHeight / 2) --
                            +(0, 0);

                        \shade[shading=axis, left color=\leftBackground, right color=\centerBackground, shading angle=90] (0, \boxHeight) -- +(\arrowWidth, \arrowHeight/2) -- +(\arrowLineOffset, \lineWidth / 2) -- +(\leftWidth, \lineWidth / 2) -- +(\leftWidth, -\lineWidth / 2) -- +(\arrowLineOffset, -\lineWidth / 2) -- +(\arrowWidth, -\arrowHeight / 2) -- +(0, 0);
                        \node[\leftForeground] at (\leftWidth / 2, \boxHeight) {\leftName};

                        \shade[shading=axis, left color=\centerBackground, right color=\centerBackground, shading angle=90] (\leftWidth, \boxHeight-\lineWidth/2) -- +(\centerWidth, 0) -- +(\centerWidth, \lineWidth) -- +(0, \lineWidth) -- +(0, 0);
                        \node[\centerForeground] at (\leftWidth + \centerWidth / 2, \boxHeight) {\centerName};

                        \shade[shading=axis, left color=\centerBackground, right color=\rightBackground, shading angle=90] (\leftWidth + \centerWidth + \rightWidth, \boxHeight) -- +(-\arrowWidth, \arrowHeight/2) -- +(-\arrowLineOffset, \lineWidth / 2) -- +(-\leftWidth, \lineWidth / 2) -- +(-\leftWidth, -\lineWidth / 2) -- +(-\arrowLineOffset, -\lineWidth / 2) -- +(-\arrowWidth, -\arrowHeight / 2) -- +(0, 0);
                        \node[\rightForeground] at (\leftWidth + \centerWidth + \rightWidth / 2, \boxHeight) {\rightName};
                    }
                }
            }
        }
    }
}

\tikzpic{comparisonTable}{1}{
    \definecolor{darkPurple}{RGB}{96, 25, 134}   
    \definecolor{darkBlue}{RGB}{0, 59, 111}      
    \definecolor{darkBrown}{RGB}{102, 51, 0}     
    \definecolor{darkGray}{RGB}{71, 79, 82}   
    \definecolor{darkRed}{RGB}{177, 0, 18}
    \definecolor{darkGreen}{RGB}{0, 147, 146}
    \definecolor{darkOrange}{RGB}{255, 111, 0}

    \pgfmathsetlengthmacro{\legendWidth}{2.75cm};
    \pgfmathsetlengthmacro{\leftWidth}{7.5cm};
    \pgfmathsetlengthmacro{\centerWidth}{2.5cm};
    \pgfmathsetlengthmacro{\rightWidth}{7.4cm};
    \pgfmathsetlengthmacro{\height}{8cm - 5.25cm};

    \pic  at (\legendWidth, -0.4cm) {arrowTableColumns={\leftWidth}{\centerWidth}{\rightWidth}{\height + 1.5cm}{0.35cm}{{\textbf{Flexibility}}/ /{\textbf{Performance}}}{{white/darkBlue}/{white/darkGreen}/{white/darkPurple}}};

    \node[drop shadow={shadow xshift=0.02cm, shadow yshift=-0.02cm, color=black}, inner sep=0, anchor=center] at (\legendWidth / 2, \height / 2) {
            \scriptsize
            \rowcolors{1}{darkGray!20}{darkGray!5}
            \begin{tabular}{|c|}
                \hline
                 \rowcolor{darkGray}
                 \color{white}{\textbf{Parameters}} \\
                 Technology \\
                 Die Area \\
                 Applications \\
                 Cores \\
                 INT Support\\
                 FXP Support \\
                 FP Support \\
                 Supply Voltage \\
                 Frequency \\
                 Performance \\
                 Energy Efficiency \\
                \hline

            \end{tabular}
    };

    \node[drop shadow={shadow xshift=0.02cm, shadow yshift=-0.02cm, color=black}, inner sep=0, anchor=center] at (\legendWidth + \leftWidth / 2, \height / 2) {
            \scriptsize
            \rowcolors{1}{darkGray!20}{darkGray!5}
            \begin{tabular}{|c|c|c}
                \hline
                    \rowcolor{darkGray}
                    \color{white}{\textbf{Conti et al.~\cite{ContiRPGMROEOHMJEGLBB23}}} & \color{white}{\textbf{Wang et al.~\cite{WangKMMP19}}} &\color{white}{\textbf{Feng et al.~\cite{FengCKKLMNSZNST22}}} \\
                    \qty{22}{\nano\metre} & \qty{40}{\nano\metre} & \qty{16}{\nano\metre} \\
                    \qty{18.7}{\square\milli\metre} & \qty{4.7}{\square\milli\metre} & \qty{20.1}{\square\milli\metre} \\
                    General Software & DFGs & DFGs \\
                    16 RISC-V & 16 PEs & 384 PEs \\
                    2-32 & N/A & 16 \\
                    N/A & 32 & N/A \\
                    16, 32 & N/A & 16 \\
                    0.5--\qty{0.8}{\volt} & 0.8--\qty{1.1}{\volt} & 0.84--\qty{1.29}{\volt} \\
                    \qty{420}{\mega\hertz} & \qty{753}{\mega\hertz} & \qty{955}{\mega\hertz} \\
                    \qty[per-mode=repeated-symbol]{6.9}{\giga\flops} & \qty[per-mode=repeated-symbol]{5.38}{\giga\ops} & \qty[per-mode=repeated-symbol]{367}{\giga\ops} \\
                    \qty[per-mode=repeated-symbol]{207}{\giga\flops\per\watt} & \qty[per-mode=repeated-symbol]{26.4}{\giga\ops\per\watt} & \qty[per-mode=repeated-symbol]{538}{\giga\ops\per\watt} \\
                \hline

            \end{tabular}
    };
    \node[drop shadow={shadow xshift=0.02cm, shadow yshift=-0.02cm, color=black}, inner sep=0, anchor=center] at (\legendWidth + \leftWidth + \centerWidth / 2, \height / 2) {
            \scriptsize
            \rowcolors{1}{darkGray!20}{darkGray!5}
            \begin{tabular}{|c|}
                \hline
                 \rowcolor{darkGray}
                 \color{white}{\textbf{This Work}} \\
                 \qty{22}{\nano\metre} \\
                 \qty{10}{\square\milli\metre} \\
                 Loop Nests \\
                 64 PEs \\
                 32 \\
                 N/A \\
                 8, 32 \\
                 0.6--\qty{1.2}{\volt} \\
                 \qty{700}{\mega\hertz} \\
                 \qty[per-mode=repeated-symbol]{537.60}{\giga\flops} \\
                 \qty[per-mode=repeated-symbol]{270}{\giga\flops\per\watt} \\
                \hline

            \end{tabular}
    };
    \node[drop shadow={shadow xshift=0.02cm, shadow yshift=-0.02cm, color=black}, inner sep=0, anchor=center] at (\legendWidth + \leftWidth + \centerWidth + \rightWidth / 2, \height / 2) {
            \scriptsize
            \rowcolors{1}{darkGray!20}{darkGray!5}
            \begin{tabular}{|c|c|c}
                \hline
                 \rowcolor{darkGray}
                 \color{white}{\textbf{Du et al.~\cite{DuTCLCLCY23}}} & \color{white}{\textbf{Moon et al.~\cite{MoonMSS23}}} &\color{white}{\textbf{Huang et al.~\cite{HuangWHKLJHCCLLTHCCC23}}} \\
                    \qty{28}{\nano\metre} & \qty{28}{\nano\metre} & \qty{22}{\nano\metre} \\
                    \qty{9.08}{\square\milli\metre} & \qty{9.61}{\square\milli\metre} & \qty{30.6}{\square\milli\metre} \\
                    Deep Learning & Deep Learning & CNN \\
                    1024 PEs & 32 PEs & N/A \\
                    8 & 1-8 & 1-8 \\
                    N/A & N/A & N/A \\
                    N/A & N/A & N/A \\
                    0.66--\qty{1.3}{\volt} & 0.46--\qty{1.13}{\volt} & 0.7--\qty{0.8}{\volt} \\
                    \qty{500}{\mega\hertz} & \qty{400}{\mega\hertz} & \qty{200}{\mega\hertz} \\
                    \qty[per-mode=repeated-symbol]{512}{\giga\ops} & \qty[per-mode=repeated-symbol]{18.9}{\tera\ops} & \qty[per-mode=repeated-symbol]{24.88}{\tera\ops} \\
                    \qty[per-mode=repeated-symbol]{11.2}{\tera\ops\per\watt} & \qty[per-mode=repeated-symbol]{127.8}{\tera\ops\per\watt} & \qty[per-mode=repeated-symbol]{251}{\tera\ops\per\watt} \\
                \hline

            \end{tabular}
    };
}

\tikzpic{overheadTable}{1}{
    \definecolor{darkPurple}{RGB}{96, 25, 134}   
    \definecolor{darkBlue}{RGB}{0, 59, 111}      
    \definecolor{darkBrown}{RGB}{102, 51, 0}     
    \definecolor{darkGray}{RGB}{71, 79, 82}   
    \definecolor{darkRed}{RGB}{177, 0, 18}
    \definecolor{darkGreen}{RGB}{0, 147, 146}
    \definecolor{darkOrange}{RGB}{255, 111, 0}

    \pgfmathsetlengthmacro{\legendWidth}{3cm};
    \pgfmathsetlengthmacro{\leftWidth}{16cm};
    \pgfmathsetlengthmacro{\height}{8cm - 3cm};
    \pgfmathsetlengthmacro{\subheight}{\height / 4};

    \node[drop shadow={shadow xshift=0.02cm, shadow yshift=-0.02cm, color=black}, inner sep=0, anchor=center] at (\legendWidth / 2, \height / 2) {
            \scriptsize
            \rowcolors{1}{darkGray!20}{darkGray!5}
            \begin{tabular}{|c|}
                \hline
                 \rowcolor{darkGray}
                 \color{white}{\textbf{Benchmark}} \\
                 Data-flow Operations \\
                 Executed Instructions \\
                 Control Overhead \\
                \hline
            \end{tabular}
    };

    \node[drop shadow={shadow xshift=0.02cm, shadow yshift=-0.02cm, color=black}, inner sep=0, anchor=center, fill=white] at (\legendWidth + \leftWidth / 2, \height / 2) {
            \scriptsize
            \rowcolors{1}{darkGray!20}{darkGray!5}
            \begin{tabular}{|c|c|c|c|c|c|c|}
                \hline
                \rowcolor{darkGray}
                \color{white}{\textbf{trmm}}         &  \color{white}{\textbf{trisolv}}         &  \color{white}{\textbf{syrk}}       &  \color{white}{\textbf{syr2k}}      &  \color{white}{\textbf{symm}}         &  \color{white}{\textbf{seidel-2d}}  &  \color{white}{\textbf{mvt}}         \\
                \qty{9648000}{}                      &  \qty{160400}{}                          &  \qty{17308800}{}                   &  \qty{69177600}{}                   &  \qty{24288000}{}                     &  \qty{142563600}{}                  &  \qty{640000}{}                      \\
                \qty{43465540}{}                     &  \qty{643217}{}                          &  \qty{69795026}{}                   &  \qty{196476513}{}                  &  \qty{77425348}{}                     &  \qty{262561052}{}                  &  \qty{2724024}{}                     \\
                \qty{77.80}{\percent}                &  \qty{75.06}{\percent}                   &  \qty{75.20}{\percent}              &  \qty{64.79}{\percent}              &  \qty{68.63}{\percent}                &  \qty{45.70}{\percent}              &  \qty{76.51}{\percent}               \\

                \hline\hline                                                                                                                                                                                                                                                                     
                \rowcolor{darkGray}                                                                                                                                                                                                                                                              
                \color{white}{\textbf{ludcmp}}       &  \color{white}{\textbf{lu}}              &  \color{white}{\textbf{jacobi-2d}}  &  \color{white}{\textbf{jacobi-1d}}  &  \color{white}{\textbf{gramschmidt}}  &  \color{white}{\textbf{gesummv}}    &  \color{white}{\textbf{gemver}}      \\
                \qty{64400400}{}                     &  \qty{32080000}{}                        &  \qty{37200600}{}                   &  \qty{159600}{}                     &  \qty{46248000}{}                     &  \qty{250750}{}                     &  \qty{1600400}{}                     \\
                \qty{109239230}{}                    &  \qty{213417020}{}                       &  \qty{102052753}{}                  &  \qty{419420}{}                     &  \qty{109437634}{}                    &  \qty{1003269}{}                    &  \qty{5289238}{}                     \\
                \qty{41.05}{\percent}                &  \qty{84.97}{\percent}                   &  \qty{63.55}{\percent}              &  \qty{61.95}{\percent}              &  \qty{57.74}{\percent}                &  \qty{75.01}{\percent}              &  \qty{69.74}{\percent}               \\

                \hline\hline                                                                                                                                                                                                                                                                     
                \rowcolor{darkGray}                                                                                                                                                                                                                                                              
                \color{white}{\textbf{gemm}}         &  \color{white}{\textbf{floyd-warshall}}  &  \color{white}{\textbf{fdtd-2d}}    &  \color{white}{\textbf{dynprog}}    &  \color{white}{\textbf{durbin}}       &  \color{white}{\textbf{doitgen}}    &  \color{white}{\textbf{covariance}}  \\
                \qty{31724000}{}                     &  \qty{500000000}{}                       &  \qty{52824000}{}                   &  \qty{4967600}{}                    &  \qty{562000}{}                       &  \qty{14620000}{}                   &  \qty{30130102}{}                    \\
                \qty{95393625}{}                     &  \qty{1251503521}{}                      &  \qty{182096857}{}                  &  \qty{23257050}{}                   &  \qty{1452037}{}                      &  \qty{59078389}{}                   &  \qty{68659335}{}                    \\
                \qty{66.74}{\percent}                &  \qty{60.05}{\percent}                   &  \qty{70.99}{\percent}              &  \qty{78.64}{\percent}              &  \qty{61.30}{\percent}                &  \qty{75.25}{\percent}              &  \qty{56.12}{\percent}               \\

                \hline\hline                                                                                                                                                                                                                                                                     
                \rowcolor{darkGray}                                                                                                                                                                                                                                                              
                \color{white}{\textbf{correlation}}  &  \color{white}{\textbf{cholesky}}        &  \color{white}{\textbf{bigc}}       &  \color{white}{\textbf{atax}}       &  \color{white}{\textbf{adi}}          &  \color{white}{\textbf{3mm}}        &  \color{white}{\textbf{2mm}}         \\
                \qty{21736080}{}                     &  \qty{32241200}{}                        &  \qty{640421}{}                     &  \qty{640400}{}                     &  \qty{72040000}{}                     &  \qty{45711900}{}                   &  \qty{36667800}{}                    \\
                \qty{61389731}{}                     &  \qty{85656422}{}                        &  \qty{2562114}{}                    &  \qty{2721501}{}                    &  \qty{227240349}{}                    &  \qty{183299647}{}                  &  \qty{132684516}{}                   \\
                \qty{64.59}{\percent}                &  \qty{62.36}{\percent}                   &  \qty{75.00}{\percent}              &  \qty{76.47}{\percent}              &  \qty{68.30}{\percent}                &  \qty{75.06}{\percent}              &  \qty{72.36}{\percent}               \\
                \hline                                                                                                                                                                                                                                                                           
            \end{tabular}
    };
}

\tikzpic{waveformPic}{9} {
    \foreach \name [count=\i] in {#5} {
       \node[anchor=east] at (0, 0 - \i * #3 + #2 / 2) {\small \name};
    }
    
    \definecolor{backgroundColor}{RGB}{255, 255, 255};
    \definecolor{slotColor}{RGB}{240, 240, 240};

    \foreach \i in {1,...,#6} {
        \foreach \j in {0,...,#7} {
            \fill[fill=backgroundColor] (0 + \j * #1, 0 - \i * #3 - #3 * 0.5 + #2 * 0.5) -- +(#1, 0) -- +(#1, #3) -- +(0, #3) circle;

            \fill[fill=slotColor] (0 + \j * #1, 0 - \i * #3 - #4) -- +(#1, 0) -- +(#1, #2 + #4 + #4) -- +(0, #2 + #4 + #4) circle;
        }
    }
    \foreach \mode\start\end\label\col\pos [count=\i] in {#9} {
        \ifthenelse{\equal{\mode}{r}}
        {
            \draw[-, very thick, draw=\col] 
                (#1 + \start * #1, 0 + #2 * \pos - #2) --
                (#1 + \start * #1, 0 + #2 * 0.5 + #2 * \pos - #2) --
                node[midway, anchor=south, above, \col] {\label}
                (#1 + \end * #1, 0 + #2 * 0.5 + #2 * \pos - #2) --
                (#1 + \end * #1, 0 + #2 * \pos - #2);
            
            \draw[-, draw=\col, dashed] 
                (#1 + \start * #1, 0 + #2 * \pos - #2) --
                (#1 + \start * #1, 0 - #6 * #3 - #3 * 0.5 + #2 * 0.5);
            \draw[-, draw=\col, dashed] 
                (#1 + \end * #1, 0 + #2 * \pos - #2) --
                (#1 + \end * #1, 0 - #6 * #3 - #3 * 0.5 + #2 * 0.5);

        }
        {
            \ifthenelse{\equal{\mode}{p}}
            {
                \draw[-, very thick, draw=\col] (#1 + \start * #1, 0 - #6 * #3 - #3 * 0.5 + #2 * 0.5) -- node[anchor=north, \col] {\label} +(0, -#4);
            }
            {
                \draw[-, very thick, draw=\col] (#1 + \start * #1, 0 - #6 * #3 - #3 * 0.5 + #2 * 0.5 - #4) -- node[at start, anchor=north,  \col] {\label} +(0, #6 * #3 + #4);
            }
        }
    }
    \foreach \data\mode [count=\i] in {#8} {
        \ifthenelse{\equal{\mode}{s}}
        {
            \foreach \point [count=\j, remember=\point as \lastPoint] in \data {
                \ifthenelse{\equal{\point}{0}}
                {
                    \ifthenelse{\equal{\lastPoint}{0}}
                    {
                        \draw[-, thick] (0 + \j * #1 - #1, 0 - \i * #3) -- +(#1, 0);
                    }
                    {
                        \draw[-, thick] (0 + \j * #1 - #1, 0 - \i * #3 + #2) -- +(0, -#2) -- +(#1, -#2);
                    }
                }
                {
                    \ifthenelse{\equal{\lastPoint}{0}}
                    {
                        \draw[-, thick] (0 + \j * #1 - #1, 0 - \i * #3) -- +(0, #2) -- +(#1, #2);
                    }
                    {
                        \draw[-, thick] (0 + \j * #1 - #1, 0 - \i * #3 + #2) -- +(#1, 0);
                    }
                }
            }
        }{}
        \ifthenelse{\equal{\mode}{b}}
        {
            \newcounter{start}
            \setcounter{start}{0}
            \foreach \trace [count=\j, remember=\trace as \lastTrace] in \data {
                \foreach \duration\name\color in \trace {
                    \node[inner sep=0, anchor=south west, minimum width=\duration * #1, minimum height=#2] at (\value{start} * #1, -\i * #3) {\name};
                    \draw[-, thick] (0 + \value{start} * #1, 0 - \i * #3 + #2) -- +(0.5 * #4, -#2) -- +(\duration * #1, -#2);
                    \draw[-, thick] (0 + \value{start} * #1, 0 - \i * #3) -- +(0.5 * #4, #2) -- +(\duration * #1, #2);

                    \addtocounter{start}{\duration}
                }
            }
        }{}
        \ifthenelse{\equal{\mode}{t}}
        {
            \foreach \trace [count=\j] in \data {
                \foreach \start\duration\name\foreground\background in \trace {
                    \colorlet{leftColor}{\background}
                    \colorlet{rightColor}{leftColor!80}
                    \node[drop shadow={shadow xshift=0.02cm, shadow yshift=-0.02cm, color=black}, shading=axis, left color=leftColor, right color=rightColor, shading angle=0, inner sep=0, anchor=south west, minimum width=\duration * #1, minimum height=#2, draw, text=\foreground] at (\start * #1, -\i * #3) {\small \textbf{\name}};
                }
            }
        }{}
    }
    \draw[blockArrowTo, very thick] (0, 0) --  (0, 0 - #6 * #3 - #3 * 0.5 + #2 * 0.5) -- node[anchor=center, midway, yshift=-#3/2] {Time} +(#7 * #1 + 2 * #1, 0);

}

\tikzpic{plotPic}{5} {
    \definecolor{darkBlue}{RGB}{0, 59, 111}      
    \pgfplotstableread{#1}{\data}
    \foreach \xLabel\xticks\xOffset in {#4} {
        \foreach \yLabel\yticks\yOffset in {#5} {
            \begin{axis}[
                width=#2,
                height=#3,
                scale only axis=true,
                xlabel = \xLabel,
                ylabel = \yLabel,
                axis lines=left, 
                thick, 
                xtick=\xticks,
                ytick=\yticks,
                xlabel style={yshift=\xOffset},
                ylabel style={yshift=-\yOffset},
                grid=both,
            ]
                \addplot[darkBlue, smooth, very thick, mark=*] table {\data};
            \end{axis}
        }
    }
}

\tikzpic{interConnectDataPlot}{2} {
    \tikzset{every node/.append style={font=\fontsize{10}{12}\selectfont}}

    \pic (array) at (0, 0) {peArrayNoBuffer=
        {#1}
        {#2}
    };
    \definecolor{darkRed}{RGB}{177, 0, 18}

    \foreach \x in {1,...,#1} {
        \foreach \y in {2,...,#2} {
            \pgfmathsetlengthmacro{\peX}{1 * (\x - 1) * 0.65cm + 0.25cm};
            \pgfmathsetlengthmacro{\peY}{1 * (\y - 1) * 0.65cm + 0.25cm};

            \ifthenelse{\equal{\x}{#1}}{}{
                \draw[thick, -{Latex[length=0.07cm]}, darkRed] (\peX, \peY) -- +(0.25cm + 0.15cm, 0);
                \draw[thick, -{Latex[length=0.07cm]}, darkRed] (\peX, \peY) -- +(-0.25cm - 0.15cm, 0);

                \draw[thick, -{Latex[length=0.07cm]}, darkRed] (\peX, \peY) -- +(0, 0.25cm + 0.15cm);
                \draw[thick, -{Latex[length=0.07cm]}, darkRed] (\peX, \peY) -- +(0, -0.25cm - 0.15cm);
            }
        }
    }
}
\tikzpic{interConnectControlPlot}{2} {
    \tikzset{every node/.append style={font=\fontsize{10}{12}\selectfont}}

    \pic (array) at (0, 0) {peArrayNoBuffer=
        {#1}
        {#2}
    };
    \definecolor{darkRed}{RGB}{177, 0, 18}

    \pic (gc) at (-3 * 0.65cm, {(#2 - 2) * 0.65cm}) {blockPic={GC}{1.15cm}{1.15cm}{/1//}{/Out//}};

    \pgfmathsetlengthmacro{\upperLeftX}{0.25cm};
    \pgfmathsetlengthmacro{\upperLeftY}{#2 * 0.65cm - 0.65cm + 0.25cm};

    \pgfmathsetlengthmacro{\lowerRightX}{#1 * 0.65cm - 0.65cm};
    \pgfmathsetlengthmacro{\lowerRightY}{0.5cm};

    \draw[blockArrowTo, thick, darkRed] (gcOut) -- (\upperLeftX, \upperLeftY);
    \draw[blockArrowTo, thick, darkRed] (\upperLeftX, \upperLeftY) -- (\upperLeftX, \lowerRightY);

    \foreach \i in {1,...,#2} {
        \ifthenelse{\equal{\i}{#2}}{}{
            \draw[blockArrowTo, thick, darkRed] (\upperLeftX, {\upperLeftY - (\i - 1) * 0.65cm}) -- (\lowerRightX, {\upperLeftY - (\i - 1) * 0.65cm});
        }
    }
}

\tikzpic{peProgram}{2} {
    \pgfmathsetlengthmacro{\columnWidth}{0.5cm};
    \pgfmathsetlengthmacro{\columnHeight}{2cm};

    \pic at (0 * \columnWidth, 0) {rotatedBlockPic={Opcode}{\columnWidth}{\columnHeight}{///}{///}};
    \pic at (1 * \columnWidth, 0) {rotatedBlockPic={Destination}{\columnWidth}{\columnHeight}{///}{///}};
    \pic at (2 * \columnWidth, 0) {rotatedBlockPic={Source}{\columnWidth}{\columnHeight}{///}{///}};
    \pic at (3 * \columnWidth, 0) {rotatedBlockPic={Source}{\columnWidth}{\columnHeight}{///}{///}};
    \pic at (4 * \columnWidth, 0) {rotatedBlockPic={Opcode}{\columnWidth}{\columnHeight}{///}{///}};
    \pic at (5 * \columnWidth, 0) {rotatedBlockPic={Wait}{\columnWidth}{\columnHeight}{///}{///}};
    \pic at (6 * \columnWidth, 0) {rotatedBlockPic={Source}{\columnWidth}{\columnHeight}{///}{///}};
    \pic at (7 * \columnWidth, 0) {rotatedBlockPic={Target}{\columnWidth}{\columnHeight}{///}{///}};
    \pic at (8 * \columnWidth, 0) {rotatedBlockPic={Target}{\columnWidth}{\columnHeight}{///}{///}};

    \pic at (0, -\columnWidth) {blockPic={Data}{\columnWidth * 4}{\columnWidth}{///}{///}};
    \pic at (4 * \columnWidth, -\columnWidth) {blockPic={Control}{\columnWidth * 5}{\columnWidth}{///}{///}};

}

\newcounter{numberOfPieces}
\newcounter{currentAngle}

\tikzpic{pieChart}{7} {
    \setcounter{numberOfPieces}{0}
    \foreach \data in {#7} {
        \foreach \name\color\fraction\val in \data {
            \addtocounter{numberOfPieces}{1}
        }
    }

    \pgfmathsetlengthmacro{\outerRadius}{#1};
    \pgfmathsetlengthmacro{\innerRadius}{#2};

    \pgfmathsetlengthmacro{\rowHeight}{#3};
    \pgfmathsetlengthmacro{\rowMargin}{#4};
    \pgfmathsetlengthmacro{\legendOffset}{#5};

    \pgfmathsetlengthmacro{\legendHeight}{\value{numberOfPieces} * (\rowHeight + \rowMargin) - \rowMargin};

    \pgfmathsetlengthmacro{\topOffset}{(2 * \outerRadius - \legendHeight) / 2};

    \foreach [count=\i] \data in {#7} {
        \foreach \name\color\fraction\val in \data {
            \pgfmathsetlengthmacro{\currentLine}{{\topOffset + (\i - 1) * (\rowHeight + \rowMargin)}};
            \node[inner sep = 0, anchor=east, minimum height=\rowHeight, align=center] at (-\legendOffset, \currentLine + \rowHeight/2) {\name};

            \colorlet{leftColor}{\color}
            \colorlet{rightColor}{leftColor!70}

            \draw[shading=axis, left color=leftColor, right color=rightColor, shading angle=0] (-\legendOffset+\rowHeight, \currentLine+\rowHeight/2) circle (\rowHeight/2);

            \ifthenelse{\equal{\val}{}}
            {
                \node[inner sep = 0, anchor=south west, minimum height=\rowHeight, align=center] at (-\legendOffset+2*\rowHeight, \currentLine) {\pgfmathparse{\fraction*100}\pgfmathprintnumber{\pgfmathresult}\,\%};
            }{
                \node[inner sep = 0, anchor=south west, minimum height=\rowHeight, align=center] at (-\legendOffset+2*\rowHeight, \currentLine) {\val~\pgfmathparse{\fraction*100}(\pgfmathprintnumber{\pgfmathresult}\,\%)};
            }
        }
    }

    \node[circle, minimum width=\outerRadius * 2, inner sep=0, align=center] (outerCircle) at (\outerRadius, \outerRadius) {#6};
    \node[circle, minimum width=\innerRadius * 2, inner sep=0] (innerCircle) at (\outerRadius, \outerRadius) {};

    \draw [drop shadow={shadow xshift=0.02cm, shadow yshift=-0.02cm, color=black}] (innerCircle.0) -- (outerCircle.0) arc[start angle=0, end angle=360,radius=\outerRadius] -- (innerCircle.360) arc[start angle=360, end angle=0,radius=\innerRadius];

    \setcounter{currentAngle}{0}

    \foreach [count=\i] \data in {#7} {
        \foreach \name\color\fraction\val in \data {
            \pgfmathsetlengthmacro{\startAngle}{\value{currentAngle}};

            \pgfmathsetlengthmacro{\angleFraction}{\fraction * 360};
            \pgfmathsetlengthmacro{\conversion}{\angleFraction};
            \addtocounter{currentAngle}{\conversion}

            \pgfmathsetlengthmacro{\endAngle}{\value{currentAngle}};
            \ifthenelse{\equal{\value{numberOfPieces}}{\i}}
            {
                \pgfmathsetlengthmacro{\endAngle}{360};
            }

            \colorlet{leftColor}{\color}
            \colorlet{rightColor}{leftColor!70}

            \draw [shading=axis, left color=leftColor, right color=rightColor, shading angle=0] (innerCircle.{\startAngle}) -- (outerCircle.{\startAngle}) arc[start angle={\startAngle}, end angle={\endAngle},radius=\outerRadius] -- (innerCircle.{\endAngle}) arc[start angle={\endAngle}, end angle={\startAngle},radius=\innerRadius];

        }
    }

}

\tikzpic{hmp4040}{0}{
    \tikzset{every node/.append style={font=\fontsize{8}{12}\selectfont}}
    \pgfmathsetlengthmacro{\connectorSize}{0.075cm};
    \pgfmathsetlengthmacro{\width}{4cm};
    \pgfmathsetlengthmacro{\height}{1.75cm};

    \pgfmathsetlengthmacro{\botHeight}{\height/175*50};
    \pgfmathsetlengthmacro{\topHeight}{\height - \botHeight};
    \pgfmathsetlengthmacro{\margins}{\width / 16};

    \pgfmathsetlengthmacro{\slotWidth}{\width / 4};

    \pgfmathsetlengthmacro{\displayWidth}{7 * \margins};
    \pgfmathsetlengthmacro{\displayHeight}{3 * \margins};
    \pgfmathsetlengthmacro{\displayRowHeight}{\displayHeight / 4};
    \pgfmathsetlengthmacro{\displayColumnWidth}{\displayWidth / 3};

    \pgfmathsetlengthmacro{\displayX}{\margins};
    \pgfmathsetlengthmacro{\displayY}{\botHeight + \margins};

    \draw[fill=lightgray] (0, 0) -- +(\width, 0) -- +(\width, \botHeight) -- +(0, \botHeight) -- +(0, 0);
    \draw[fill=gray] (0, \botHeight) -- +(\width, 0) -- +(\width, \topHeight) -- +(0, \topHeight) -- +(0, 0);

    \draw[fill=white] (\displayX, \displayY) -- + (\displayWidth, 0) -- +(\displayWidth, \displayHeight) -- +(0, \displayHeight) -- +(0, 0);

    \foreach \i in {1,...,3} {
        \draw[very thin, -] (\displayX, \displayY + \i * \displayRowHeight) -- +(\displayWidth, 0);
    }



    \pgfmathsetlengthmacro{\displayFontSize}{\displayRowHeight * 0.72};

    \node[font=\fontsize{\displayFontSize}{8}\selectfont] at (\displayX + \displayColumnWidth / 2, \displayY + \displayRowHeight / 2 + 2 * \displayRowHeight) {\qty{0.9}{\volt}};
    \node[font=\fontsize{\displayFontSize}{8}\selectfont] at (\displayX + \displayColumnWidth / 2 + 1 * \displayColumnWidth, \displayY + \displayRowHeight / 2 + 2 * \displayRowHeight) {\qty{2.74}{\watt}};
    \node[font=\fontsize{\displayFontSize}{8}\selectfont] at (\displayX + \displayColumnWidth / 2 + 2 * \displayColumnWidth, \displayY + \displayRowHeight / 2 + 2 * \displayRowHeight) {\qty{3.0}{\ampere}};

    \node[font=\fontsize{\displayFontSize}{8}\selectfont] at (\displayX + \displayColumnWidth / 2, \displayY + \displayRowHeight / 2 + 3 * \displayRowHeight) {\qty{12.2}{\volt}};
    \node[font=\fontsize{\displayFontSize}{8}\selectfont] at (\displayX + \displayColumnWidth / 2 + 1 * \displayColumnWidth, \displayY + \displayRowHeight / 2 + 3 * \displayRowHeight) {\qty{10.62}{\watt}};
    \node[font=\fontsize{\displayFontSize}{8}\selectfont] at (\displayX + \displayColumnWidth / 2 + 2 * \displayColumnWidth, \displayY + \displayRowHeight / 2 + 3 * \displayRowHeight) {\qty{870.5}{\milli\ampere}};

    \node at (0.75 * \width, 2.5 * \botHeight) {\scriptsize R\&S HMP4040};

    \definecolor{darkBlue}{RGB}{0, 59, 111}      
    \definecolor{darkGray}{RGB}{71, 79, 82}   
    \definecolor{darkRed}{RGB}{177, 0, 18}

    \foreach \i in {1,...,4} {
        \pgfmathsetlengthmacro{\centerX}{(\i - 1) * \slotWidth + \slotWidth / 2};
        \pgfmathsetlengthmacro{\centerY}{\botHeight / 2};

        \draw[fill=darkGray] (\centerX-1.5*\connectorSize, \centerY+1.5*\connectorSize) circle (\connectorSize);
        \draw[fill=darkGray] (\centerX+1.5*\connectorSize, \centerY+1.5*\connectorSize) circle (\connectorSize);

        \draw[fill=darkBlue] (\centerX-1.5*\connectorSize, \centerY-1.5*\connectorSize) circle (\connectorSize);
        \draw[fill=darkRed] (\centerX+1.5*\connectorSize, \centerY-1.5*\connectorSize) circle (\connectorSize);

        \draw[fill=black] (\centerX-1.5*\connectorSize, \centerY+1.5*\connectorSize) circle (\connectorSize / 2);
        \draw[fill=black] (\centerX+1.5*\connectorSize, \centerY+1.5*\connectorSize) circle (\connectorSize / 2);

        \draw[fill=blue] (\centerX-1.5*\connectorSize, \centerY-1.5*\connectorSize) circle (\connectorSize / 2);
        \draw[fill=red] (\centerX+1.5*\connectorSize, \centerY-1.5*\connectorSize) circle (\connectorSize / 2);

        \node[inner sep=0] (SenseMinusSlot\i) at (\centerX-1.5*\connectorSize, \centerY+1.5*\connectorSize) {};
        \node[inner sep=0] (SensePlusSlot\i) at (\centerX+1.5*\connectorSize, \centerY+1.5*\connectorSize) {};

        \node[inner sep=0] (DriveMinusSlot\i) at (\centerX-1.5*\connectorSize, \centerY-1.5*\connectorSize) {};
        \node[inner sep=0] (DrivePlusSlot\i) at (\centerX+1.5*\connectorSize, \centerY-1.5*\connectorSize) {};
    }
}

\tikzpic{testSetup}{0}{
    \tikzset{every node/.append style={font=\fontsize{8}{12}\selectfont}}
    \pic (hmp) at (0, 0) {hmp4040={}};

    \node[draw, minimum width=2cm, minimum height=1.5cm, inner sep=0, align=center, anchor=south west] at (1, -2.5) {\footnotesize Xilinx\\\footnotesize ZCU104};
    \node[draw, minimum width=1.25cm, minimum height=0.6cm, inner sep=0, align=center, anchor=south west] at (2.8, -2.0) {\footnotesize Chip};
    \node[draw, minimum width=0.6cm, minimum height=0.2cm, inner sep=0, align=center, anchor=north west, rotate=90] at (2.8, -2.0) {\tiny FMC};

    \coordinate (fpgaMinus)at (1, -2.5cm + 1.5cm/2 - 0.125cm);
    \coordinate (fpgaPlus)at (1, -2.5cm + 1.5cm/2 + 0.125cm);

    \coordinate (coreSenseMinus) at (2.8cm+1.25cm/2 - 0.2cm, -1.4cm);
    \coordinate (coreDriveMinus) at (2.8cm+1.25cm/2 - 0.1cm, -1.4cm);
    \coordinate (coreDrivePlus) at (2.8cm+1.25cm/2 + 0.1cm, -1.4cm);
    \coordinate (coreSensePlus) at (2.8cm+1.25cm/2 + 0.2cm, -1.4cm);

    \draw[-,thick, blue] (hmpDriveMinusSlot1) |- (fpgaMinus);
    \draw[-,thick, red] (hmpDrivePlusSlot1) |- (fpgaPlus);

    \draw[-,thick, blue] (hmpDriveMinusSlot2) -- +(0, -0.7cm) -| (coreDriveMinus);
    \draw[-,thick, red] (hmpDrivePlusSlot2) -- +(0, -0.5cm) -| (coreDrivePlus);

    \draw[-,thick, black] (hmpSenseMinusSlot2) -| +(-0.2cm, -1.1cm) -| (coreSenseMinus);
    \draw[-,thick, black] (hmpSensePlusSlot2) -| +(0.2cm, -0.6cm) -| (coreSensePlus);

    \node[anchor=north east] at ([xshift=-0cm]fpgaMinus) {\tiny \qty{12.2}{\volt}};

    \node[anchor=south west, align=center] at ([yshift=0.2cm]coreSensePlus) {\tiny \qty{0.9}{\volt}};
    \node[anchor=south west, align=center] at ([xshift=-0cm]coreSensePlus) {\tiny +Sense};
}

\tikzpic{fancyArrow}{4}{
    \pgfmathsetlengthmacro{\leftWidth}{#1};
    \pgfmathsetlengthmacro{\height}{#2};
    \pgfmathsetlengthmacro{\lineWidth}{#3};

    \pgfmathsetlengthmacro{\arrowHeight}{\lineWidth * 2.00};
    \pgfmathsetlengthmacro{\arrowWidth}{\arrowHeight * 0.75};
    \pgfmathsetlengthmacro{\arrowLineOffset}{\arrowWidth / 3 + ((\arrowWidth / 3 * 2) / (\arrowHeight / 2)) * (\lineWidth / 2)}

    \pgfmathsetlengthmacro{\boxHeight}{\height-0.5 * \arrowHeight};

    \foreach \leftColor\rightColor in {#4} {
        \draw[drop shadow={shadow xshift=0.02cm, shadow yshift=-0.02cm, color=black}] 
            (0, 0) -- 
            +(\arrowWidth, \arrowHeight/2) -- 
            +(\arrowLineOffset, \lineWidth / 2) -- 
            +(\leftWidth - \arrowLineOffset, \lineWidth / 2) -- 
            +(\leftWidth - \arrowLineOffset, \lineWidth / 2) -- 
            +(\leftWidth - \arrowLineOffset, -\lineWidth / 2) -- 
            +(\arrowLineOffset, -\lineWidth / 2) -- 
            +(\arrowWidth, -\arrowHeight / 2) --
            +(0, 0);

        \shade[shading=axis, left color=\leftColor, right color=\rightColor, shading angle=00] 
            (0, 0) -- 
            +(\arrowWidth, \arrowHeight/2) -- 
            +(\arrowLineOffset, \lineWidth / 2) -- 
            +(\leftWidth - \arrowLineOffset, \lineWidth / 2) -- 
            +(\leftWidth - \arrowLineOffset, \lineWidth / 2) -- 
            +(\leftWidth - \arrowLineOffset, -\lineWidth / 2) -- 
            +(\arrowLineOffset, -\lineWidth / 2) -- 
            +(\arrowWidth, -\arrowHeight / 2) --
            +(0, 0);
    }
}

\tikzpic{turtleFlow}{4}{
    \pgfmathsetlengthmacro{\fontHeight}{#2 / 2};
    \tikzset{every node/.append style={font=\fontsize{\fontHeight}{\fontHeight}\selectfont}}
    \pgfmathsetlengthmacro{\blockWidth}{#1};
    \pgfmathsetlengthmacro{\blockHeight}{#2};
    \pgfmathsetlengthmacro{\verticalMargin}{#3};
    \pgfmathsetlengthmacro{\horizontalMargin}{#4};

    \definecolor{darkPurple}{RGB}{96, 25, 134}   
    \definecolor{darkBlue}{RGB}{0, 59, 111}      
    \definecolor{darkBrown}{RGB}{102, 51, 0}     
    \definecolor{darkGray}{RGB}{71, 79, 82}   
    \definecolor{darkRed}{RGB}{177, 0, 18}
    \definecolor{darkGreen}{RGB}{0, 147, 39}
    \definecolor{darkCyan}{RGB}{0, 147, 146}
    \definecolor{darkOrange}{RGB}{255, 111, 0}

    \pic[rotate=90] at ({(\blockWidth + \horizontalMargin) * 2 + \blockWidth/2}, {\blockHeight * 7 + \verticalMargin * 6 + \verticalMargin / 3 * 2+\blockHeight+\verticalMargin/3*2}) {fancyArrow={\blockHeight*4}{0.3*\blockWidth}{0.15*\blockWidth}{darkGray!30/darkGray!30}};
    \pic at ({(\blockWidth + \horizontalMargin) * 1 - \horizontalMargin / 3 }, {(\blockHeight + \verticalMargin) * 8 - \verticalMargin / 3 + 1*\blockHeight+\verticalMargin/3*2})
    {grayBlockPic={}{\blockWidth * 3 + \horizontalMargin * 2 + \horizontalMargin / 3 * 2 + \blockHeight / 3 * 2}{\blockHeight * 2 + \verticalMargin * 1 + \verticalMargin / 3 * 2}{///}{///}};

    \pic at ({(\blockWidth + \horizontalMargin) * 3 + \blockWidth + \horizontalMargin / 3 }, {(\blockHeight + \verticalMargin) * 8 - \verticalMargin / 3 + 1*\blockHeight+\verticalMargin/3*2})
    {rotatedBlockPic={Project}{\blockHeight / 3 * 2}{\blockHeight * 2 + \verticalMargin * 1 + \verticalMargin / 3 * 2}{///}{///}};
    \pic at ({(\blockWidth + \horizontalMargin) * 0 - \horizontalMargin / 3 }, {(\blockHeight + \verticalMargin) * 4 - \verticalMargin / 3})
    {grayBlockPic={}{\blockWidth * 5 + \horizontalMargin * 4 + \horizontalMargin / 3 * 2 + \blockHeight / 3 * 2}{\blockHeight * 4 + \verticalMargin * 3 + \verticalMargin / 3 * 2}{///}{///}};

    \pic at ({(\blockWidth + \horizontalMargin) * 4 + \blockWidth + \horizontalMargin / 3 }, {(\blockHeight + \verticalMargin) * 4 - \verticalMargin / 3})
    {rotatedBlockPic={Mapping}{\blockHeight / 3 * 2}{\blockHeight * 4 + \verticalMargin * 3 + \verticalMargin / 3 * 2}{///}{///}};

    \pic at ({(\blockWidth + \horizontalMargin) * 2 - \horizontalMargin / 3 }, {(\blockHeight + \verticalMargin) * 1 - \verticalMargin / 3})
    {grayBlockPic={}{\blockWidth * 3 + \horizontalMargin * 2 + \horizontalMargin / 3 * 2 + \blockHeight / 3 * 2}{\blockHeight * 3 + \verticalMargin * 2 + \verticalMargin / 3 * 2}{///}{///}};

    \pic at ({(\blockWidth + \horizontalMargin) * 4 + \blockWidth + \horizontalMargin / 3 }, {(\blockHeight + \verticalMargin) * 1 - \verticalMargin / 3})
    {rotatedBlockPic={Integration}{\blockHeight / 3 * 2}{\blockHeight * 3 + \verticalMargin * 2 + \verticalMargin / 3 * 2}{///}{///}};

    \pic at ({(\blockWidth + \horizontalMargin) * 0 - \horizontalMargin / 3 }, {(\blockHeight + \verticalMargin) * 0 - \verticalMargin / 3 - \blockHeight*1})
    {grayBlockPic={}{\blockWidth * 5 + \horizontalMargin * 3 + \horizontalMargin / 3 * 2 + \blockHeight / 3 * 2}{\blockHeight * 1 + \verticalMargin * 0 + \verticalMargin / 3 * 2}{///}{///}};

    \pic at ({(\blockWidth + \horizontalMargin) * 4 + \blockWidth + \horizontalMargin / 3 }, {(\blockHeight + \verticalMargin) * 0 - \verticalMargin / 3 - \blockHeight*1})
    {rotatedBlockPic={Target}{\blockHeight / 3 * 2}{\blockHeight * 1 + \verticalMargin * 0 + \verticalMargin / 3 * 2}{///}{///}};

    \pic[rotate=90] at ({(\blockWidth + \horizontalMargin) * 0 + \blockWidth/2}, 0) {fancyArrow={(\blockHeight + \verticalMargin) * 9 - \verticalMargin*3}{0.3*\blockWidth}{0.15*\blockWidth}{darkGray!30/darkGreen!90}};
    \pic[rotate=90] at ({(\blockWidth + \horizontalMargin) * 1 + \blockWidth/2}, 0) {fancyArrow={(\blockHeight + \verticalMargin) * 9 - \verticalMargin*3}{0.3*\blockWidth}{0.15*\blockWidth}{darkGray!30/darkBlue!90}};
    \pic[rotate=90] at ({(\blockWidth + \horizontalMargin) * 2 + \blockWidth/2}, 0) {fancyArrow={(\blockHeight + \verticalMargin) * 9 - \verticalMargin*3}{0.3*\blockWidth}{0.15*\blockWidth}{darkGray!30/darkPurple!90}};
    \pic[rotate=90] at ({(\blockWidth + \horizontalMargin) * 3 + \blockWidth/2}, 0) {fancyArrow={(\blockHeight + \verticalMargin) * 9 - \verticalMargin*3}{0.3*\blockWidth}{0.15*\blockWidth}{darkGray!30/darkOrange!90}};
    \pic[rotate=90] at ({(\blockWidth + \horizontalMargin) * 4 + \blockWidth/2}, 0) {fancyArrow={(\blockHeight + \verticalMargin) * 9 - \verticalMargin*3}{0.3*\blockWidth}{0.15*\blockWidth}{darkGray!30/darkRed!90}};

    \pic at ({(\blockWidth + \horizontalMargin) * 0}, -\blockHeight*1) {coloredBlockPic={\faPython~Python}{\blockWidth}{\blockHeight}{///}{///}{darkGreen/darkGreen!90}{white}};
    \pic at ({(\blockWidth + \horizontalMargin) * 1}, -\blockHeight*1) {coloredBlockPic={\faCode~SW-Simulator}{\blockWidth}{\blockHeight}{///}{///}{darkBlue/darkBlue!90}{white}};
    \pic at ({(\blockWidth + \horizontalMargin) * 2}, -\blockHeight*1) {coloredBlockPic={\faWaveSquare~RTL-Simulator}{\blockWidth}{\blockHeight}{///}{///}{darkPurple/darkPurple!90}{white}};
    \pic at ({(\blockWidth + \horizontalMargin) * 3}, -\blockHeight*1) {coloredBlockPic={\faMicrochip~FPGA}{\blockWidth}{\blockHeight}{///}{///}{darkOrange/darkOrange!90}{white}};
    \pic at ({(\blockWidth + \horizontalMargin) * 4}, -\blockHeight*1) {coloredBlockPic={\faMicrochip~ASIC}{\blockWidth}{\blockHeight}{///}{///}{darkRed/darkRed!90}{white}};

    \pic at ({(\blockWidth + \horizontalMargin) * 2}, \blockHeight + \verticalMargin) {coloredBlockPic={\faCopy[regular]~Testbenches}{\blockWidth}{\blockHeight}{///}{///}{darkGray/darkGray!95}{white}};
    \pic at ({(\blockWidth + \horizontalMargin) * 3}, \blockHeight + \verticalMargin) {coloredBlockPic={\faLinux~Driver}{\blockWidth * 2 + \horizontalMargin}{\blockHeight}{///}{///}{darkGray/darkGray!95}{white}};

    \pic at ({(\blockWidth + \horizontalMargin) * 2}, {(\blockHeight + \verticalMargin) * 2}) {blockPic={\faFile[regular]~Binary Configuration}{\blockWidth * 3 + \horizontalMargin * 2}{\blockHeight}{///}{///}};
    \pic at ({(\blockWidth + \horizontalMargin) * 2}, {(\blockHeight + \verticalMargin) * 3}) {coloredBlockPic={\faTools~Binarization}{\blockWidth * 3 + \horizontalMargin * 2}{\blockHeight}{///}{///}{darkGray/darkGray!95}{white}};
    \pic at ({(\blockWidth + \horizontalMargin) * 1}, {(\blockHeight + \verticalMargin) * 4}) {blockPic={\faFileCode[regular]~Concrete Configuration}{\blockWidth * 4 + \horizontalMargin * 3}{\blockHeight}{///}{///}};
    \pic at ({(\blockWidth + \horizontalMargin) * 1}, {(\blockHeight + \verticalMargin) * 5}) {coloredBlockPic={\faTools~Instantiation}{\blockWidth * 4 + \horizontalMargin * 3}{\blockHeight}{///}{///}{darkGray/darkGray!95}{white}};
    \pic at ({(\blockWidth + \horizontalMargin) * 0}, {(\blockHeight + \verticalMargin) * 6}) {blockPic={\faFileCode[regular]~Python Code}{\blockWidth * 1 + \horizontalMargin * 0}{\blockHeight}{///}{///}};
    \pic at ({(\blockWidth + \horizontalMargin) * 1}, {(\blockHeight + \verticalMargin) * 6}) {blockPic={\faFileCode[regular]~Symbolic Configuration}{\blockWidth * 4 + \horizontalMargin * 3}{\blockHeight}{///}{///}};
    \pic at ({(\blockWidth + \horizontalMargin) * 1}, {(\blockHeight + \verticalMargin) * 7}) {coloredBlockPic={\faTools~Compilation}{\blockWidth * 4 + \horizontalMargin * 3}{\blockHeight}{///}{///}{darkGray/darkGray!95}{white}};
    \pic at ({(\blockWidth + \horizontalMargin) * 0}, {(\blockHeight + \verticalMargin) * 7}) {coloredBlockPic={\faTools~Code-Generator}{\blockWidth}{\blockHeight}{///}{///}{darkGray/darkGray!95}{white}};

    \pic at ({(\blockWidth + \horizontalMargin) * 1.5}, {(\blockHeight + \verticalMargin) * 8 + 1*\blockHeight+\verticalMargin/3*2}) {blockPic={\faFile*[regular]~Parameters}{\blockWidth}{\blockHeight}{///}{///}};
    \pic at ({(\blockWidth + \horizontalMargin) * 2.5}, {(\blockHeight + \verticalMargin) * 8 + 1*\blockHeight+\verticalMargin/3*2}) {blockPic={\faFile*[regular]~Constraints}{\blockWidth}{\blockHeight}{///}{///}};
    \pic at ({(\blockWidth + \horizontalMargin) * 1}, {(\blockHeight + \verticalMargin) * 9 + 1*\blockHeight+\verticalMargin/3*2}) {blockPic={\faFileCode[regular]~Data-Generator}{\blockWidth}{\blockHeight}{///}{///}};
    \pic at ({(\blockWidth + \horizontalMargin) * 2}, {(\blockHeight + \verticalMargin) * 9 + 1*\blockHeight+\verticalMargin/3*2}) {blockPic={\faFileCode[regular]~PAULA}{\blockWidth}{\blockHeight}{///}{///}};
    \pic at ({(\blockWidth + \horizontalMargin) * 3}, {(\blockHeight + \verticalMargin) * 9 + 1*\blockHeight+\verticalMargin/3*2}) {blockPic={\faFileCode[regular]~Architecture}{\blockWidth}{\blockHeight}{///}{///}};



}

\tikzpic{cgraPE}{2}{
    \pgfmathsetlengthmacro{\slotWidth}{#1 / 2.75};
    \pgfmathsetlengthmacro{\fontHeight}{#2};
    \pgfmathsetlengthmacro{\arrowLength}{\slotWidth*0.10};
    \tikzset{every node/.append style={font=\fontsize{\fontHeight}{\fontHeight}\selectfont}}

    \pic (background) at (-0.5 * \slotWidth, -1.5*\slotWidth) {grayBlockPic={}{#1}{#1}{}{}};
    \coordinate (SW) at (-0.5 * \slotWidth, -1.5*\slotWidth);
    \coordinate (NW) at (-0.5 * \slotWidth, -1.5*\slotWidth+#1);

    \pic (instr) at (1 * \slotWidth, -1 * \slotWidth) {blockPic={Instruction\\Memory}{\slotWidth * 0.75}{\slotWidth * 0.75}{/3/2/2}{/{Alu, North, East}/{West, South}/{Lr, Bar}}};
    \pic (bar) at (0 * \slotWidth, 0 * \slotWidth) {blockPic={Crossbar\\Switch}{\slotWidth * 0.75}{\slotWidth * 0.75}{4/3/4/2}{{n0, n1, Alu0, Alu1}/{e0, e1, Instr}/{s0, s1, Lr, Alu2}/{w0, w1}}};

    \pic (lr) at (0 * \slotWidth, -1 * \slotWidth) {blockPic={Local\\Registers}{\slotWidth * 0.75}{\slotWidth * 0.75}{2/1//}{{Bar, Alu}/Instr//}};
    \pic (alu) at (1 * \slotWidth,  0 * \slotWidth) {aluPic={ALU}{\slotWidth * 0.75}{\slotWidth * 0.75}{}{}};

    \pic (southIO) at (-0.25 * \slotWidth,  -1.5 * \slotWidth) {blockPic={South IO}{\slotWidth * 2.25}{\slotWidth * 0.25}{1///}{p///}};
    \pic (northIO) at (-0.25 * \slotWidth,  1.0 * \slotWidth) {blockPic={North IO}{\slotWidth * 2.25}{\slotWidth * 0.25}{//1/}{//p/}};
    \pic (eastIO) at (2.0 * \slotWidth,  -1.25 * \slotWidth) {rotatedBlockPic={East IO}{\slotWidth * 0.25}{\slotWidth * 2.25}{///1}{///p}};
    \pic (westIO) at (-0.5 * \slotWidth,  -1.25 * \slotWidth) {rotatedBlockPic={West IO}{\slotWidth * 0.25}{\slotWidth * 2.25}{/1//}{/p//}};

    \draw [thick, -{Latex[length=\arrowLength]}] (barn0) -- (barn0 |- northIOp);
    \draw [thick, {Latex[length=\arrowLength]}-] (barn1) -- (barn1 |- northIOp);
    \draw [thick, -{Latex[length=\arrowLength]}] (barw0) -- (barw0 -| westIOp);
    \draw [thick, {Latex[length=\arrowLength]}-] (barw1) -- (barw1 -| westIOp);

    \draw [thick, -{Latex[length=\arrowLength]}] (bare0) -- ([xshift=\slotWidth*0.2]bare0) |- ([yshift=0.05*\slotWidth]eastIOp);
    \draw [thick, {Latex[length=\arrowLength]}-] (bare1) -- ([xshift=\slotWidth*0.25]bare1) |- ([yshift=-0.05*\slotWidth]eastIOp);

    \draw [thick, {Latex[length=\arrowLength]}-] (bars0) -- ([yshift=-\slotWidth*0.1]bars0) -| ([xshift=-1.05*\slotWidth]southIOp);
    \draw [thick, -{Latex[length=\arrowLength]}] (bars1) -- ([yshift=-\slotWidth*0.15]bars1) -| ([xshift=-0.95*\slotWidth]southIOp);

    \draw [thick, -{Latex[length=\arrowLength]}] (barAlu0) |- ([yshift=\slotWidth*0.2]aluRs0) -- (aluRs0);
    \draw [thick, -{Latex[length=\arrowLength]}] (barAlu1) |- ([yshift=\slotWidth*0.15]aluRs1) -- (aluRs1);
    \draw [thick, -{Latex[length=\arrowLength]}] (aluRd) |- ([yshift=-\slotWidth*0.125]barAlu2) -- (barAlu2);
    \draw [thick, -{Latex[length=\arrowLength]}] (aluRd) -- ([yshift=-\slotWidth*0.125]aluRd) -| (lrAlu);
    \draw [thick, -{Latex[length=\arrowLength]}] (lrBar) |- ([yshift=-\slotWidth*0.2]barLr) -| (barLr);

    \draw [dashed, -{Latex[length=\arrowLength]}] (instrAlu) -- ([xshift=\slotWidth*0.1]instrAlu) |- (aluEast0);
    \draw [dashed, -{Latex[length=\arrowLength]}] (instrLr) -- ([xshift=-\slotWidth*0.1]instrLr) |- (lrInstr);
    \draw [dashed, -{Latex[length=\arrowLength]}] (instrBar) -| ([xshift=\slotWidth*0.1]barInstr) |- (barInstr);

    \draw [dashed, -{Latex[length=\arrowLength]}] (instrEast) -- (instrEast -| eastIOp);
    \draw [dashed, -{Latex[length=\arrowLength]}] (instrSouth) -- (instrSouth |- southIOp);
    \draw [dashed, -{Latex[length=\arrowLength]}] (instrNorth) -| ([xshift=1.05*\slotWidth]northIOp);
    \draw [dashed, -{Latex[length=\arrowLength]}] (instrWest) |- ([yshift=-1.05*\slotWidth]westIOp);

}

\tikzpic{cgra}{4}{
    \pgfmathsetlengthmacro{\peSize}{#3};
    \pgfmathsetlengthmacro{\peOffset}{#4};

    \pic (background) at (-\peOffset, \peSize) {grayBlockPic={}{\peSize * #2 + \peOffset * #2 + \peOffset * 2 + \peSize}{\peSize * #2 + \peOffset * #2 + \peOffset}{2/2/2/2}{{n0,n1}/{e0,e1}/{s0,s1}/{w0, w1}}};
    \pic (spm) at (0, \peSize + \peOffset) {rotatedBlockPic={Scratchpad Memory (SPM)}{\peSize}{\peSize * #2 + \peOffset * #2 - \peOffset}{2/2/2/2}{{n0,n1}/{e0,e1}/{s0,s1}/{w0, w1}}};

    \foreach \x in {1,...,#1}{
        \foreach \y in {1,...,#2}{
            \pic (PE\x\y) at ({\x * (\peSize + \peOffset)}, {\y * (\peSize + \peOffset)}) {blockPic={PE}{\peSize}{\peSize}{2/2/2/2}{{n0,n1}/{e0,e1}/{s0,s1}/{w0, w1}}};
        }
    }
    \foreach \x in {1,...,#1}{
        \foreach \y in {1,...,#2}{
            \ifthenelse{\x = 1}{}{
                \draw[->] (PE\x\y w0) -- (PE\the\numexpr\x-1\relax\y e0);
                \draw[<-] (PE\x\y w1) -- (PE\the\numexpr\x-1\relax\y e1);
            }
            \ifthenelse{\y = 1}{}{
                \draw[->] (PE\x\y s0) -- (PE\x\the\numexpr\y-1\relax n0);
                \draw[<-] (PE\x\y s1) -- (PE\x\the\numexpr\y-1\relax n1);
            }
        }
    }
    \foreach \y in {1,...,#2}{
        \draw[->] (PE1\y w0) -- (PE1\y w0 -| spme0);
        \draw[<-] (PE1\y w1) -- (PE1\y w1 -| spme1);
    }

    \pic (cgra) at ({#1 * (\peSize + \peOffset) + \peSize + \peOffset + \peOffset}, {#2 * (\peSize + \peOffset)}) {cgraPE={\peSize * 4.5}{\peSize * 4 * 0.065}};
    \draw[fill=black, opacity=0.5] ({#1 * (\peSize + \peOffset)}, {#2 * (\peSize + \peOffset)}) -- (cgraSW) -- (cgraNW) -- ({#1 * (\peSize + \peOffset)}, {#2 * (\peSize + \peOffset)+\peSize}) -- cycle;
}

\newcommand{\eg}{e.\,g.}
\newcommand{\ie}{i.\,e.}
\newcommand*\circled[1]{\tikz[baseline=(char.base)]{
            \node[shape=circle,draw,inner sep=1pt] (char) {#1};}}
\definecolor{cmark_green}{HTML}{00A64F}
\newcommand{\cmark}{\textcolor{cmark_green}{\ding{51}}}
\definecolor{xmark_red}{HTML}{ED1B23}
\newcommand{\xmark}{\textcolor{xmark_red}{\ding{55}}}
\definecolor{mmark_yellow}{HTML}{ED7A1B}
\newcommand{\mmark}{\textcolor{mmark_yellow}{\ding{51}}}

\newcolumntype{C}[1]{>{\centering\let\newline\\\arraybackslash\hspace{0pt}}m{#1}}

\newcommand{\secRef}[1]{Section~\ref{#1}}
\newcommand{\codeRef}[1]{Listing~\ref{#1}}
\newcommand{\figRef}[1]{Figure~\ref{#1}}
\newcommand{\tableRef}[1]{Table~\ref{#1}}
\setlength\arraycolsep{1pt}

\newcommand{\factor}[1]{\qty{#1}{\times}}

\title{Evaluation of CGRA Toolchains}
\author{
	\IEEEauthorblockN{Dominik Walter, Marita Halm, Daniel Seidel, Indrayudh Ghosh, \\Christian Heidorn, Frank Hannig, J\"urgen Teich}
	\IEEEauthorblockA{cs12-alpaca@fau.de\\
	Hardware/Software Co-Design, Department of Computer Science\\
	Friedrich-Alexander-Universit\"at Erlangen-N\"urnberg (FAU), Germany
			}
}

\maketitle

\begin{abstract}
    Increasing demands for computing power also propel the need for energy-efficient SoC accelerator architectures.
    One class for such accelerators are so-called processor arrays, which typically integrate a two-dimensional mesh of interconnected processing elements~(PEs).
    Such arrays are specifically designed to accelerate the execution of multidimensional nested loops by exploiting the intrinsic parallelism of such loops.
    Coarse-grained reconfigurable arrays~(CGRAs) belong to this class of accelerator architectures.
    In this work, we analyze four toolchains for mapping loop programs onto CGRAs and compare the resulting mappings wrt. performance, \ie, latency.
    While most toolchains succeed in simpler kernels like general matrix multiplication, some struggle to find valid mappings for more complex loops like a triangular solver.
    Furthermore, we observe that the considered CGRA mappers generally tend to underutilize the available PEs.
\end{abstract}


\section{Introduction}
    Coarse-grained reconfigurable arrays~(CGRAs) were first presented in the 1990s, and since then, many CGRA architectures have been developed by both industry and academia, see, \eg,~\cite{26_CGRA_Overview,24_CGRA_Taxonomy}.
    In this study, we evaluate and compare the mapping quality of four publicly available CGRA toolchains in terms of performance, \ie, the achieved latency on a set of selected loop benchmarks.
    More specifically, we introduce the fundamentals of CGRAs and mapping of data flow graphs in Sections~\ref{sec:cgra} and \ref{sec:cgra:mapping}, followed by discussing available toolchains in \secRef{sec:tools}.
    Afterward, we evaluate the mapping tolls in terms of achieved latency in \secRef{sec:evaluation} and summarize our study in \secRef{sec:summary}.
\section{Coarse-Grained Reconfigurable Arrays}
\label{sec:cgra}
    According to~\cite{19_CGRA}, a typical CGRA architecture consists of a network of interconnected PEs arranged in a two-dimensional grid, as shown in Figure \ref{fig:mainFigCgra}~(right).
    To keep the hardware simple and modular, each PE contains one functional unit~(FU), a set of local registers potentially arranged in a register file, a crossbar switch, and an instruction memory.
    The FU usually supports different arithmetic, logic, and memory operations at the word level.
    The local registers are used as temporary data storage for intermediate results.
    The crossbar connects the PE with its adjacent neighbors, enabling data transfer between neighboring PEs in a single cycle.
    The operation performed by the FU and the routing for the crossbar can be configured at the granularity of clock cycles.
    The instruction memory can store a sequence of predetermined per-cycle configurations to execute one loop iteration.
    Most CGRAs also support conditional execution by predication, \ie, the execution of some instructions is masked by a predication bit that was set by a conditional instruction.
    According to \figRef{fig:mainFigCgra}, typically a subset of PEs have a direct access to an accompanying on-chip scratchpad memory (SPM) that can buffer input and output locally.
    Moreover, because only neighboring PEs can read or write data within one clock cycle, transferring data to a PE further away usually requires a propagation taking multiple cycles, while the intermediate PEs are then occupied for communication.
    HyCUBE~\cite{2_HyCUBE,WangKMMP19} alleviates these issues by a reconfigurable interconnect with single-cycle multi-hop connections.

\section{Data Flow Graph Mapping}
\label{sec:cgra:mapping}
        CGRAs are designed to accelerate nested loops.
        Common to mapping tools for CGRAs is a loop representation in the form of a data flow graph (DFG) that is extracted from a given C/C++ program.
        The DFG is then statically mapped and scheduled on the PEs, aiming to achieve the lowest possible \emph{initiation interval}~\II, such that each node is repeatedly executed by the same PE every \II cycles.
        Consider \figRef{fig:mainFigCgra}.
        It shows a simplified but representative DFG of a typical 3-dimensional loop nest for computing a matrix-matrix multiplication.
        Each node denotes one operation, and the edges show the dependencies between the operations.
        For the execution of each loop iteration $(i, j, k)$, four types of computations are involved:
        {(a)} Determination of the current loop indices---the corresponding operations are shown in \figRef{fig:mainFigCgra} on the left.
        Each loop index computation requires three operations, \eg, to compute the innermost loop (index $i$), a \texttt{Sel}, \texttt{Add}, and \texttt{Cmp} operation is needed.
        \texttt{Sel} is a multiplex operation that uses the result of the \texttt{Cmp} instruction to either forward the output of the \texttt{Add} operation or zero.
        The \texttt{Cmp} compares the result of the \texttt{Add} operation, which increments the current loop index, against a predefined constant, here, the loop bound.
        Note that the data dependencies towards the \texttt{Sel} operations are inter-iteration dependencies.
        This effectively implements a cyclic accumulator.
        Furthermore, the result of the \texttt{Cmp} operation can also be used as an addend for the \texttt{Add} operation of the next loop index.
        Therefore, the second level (index $j$) is only incremented once the first one reaches its loop bound, implementing a two-dimensional loop counter.
        This can be repeated for a third outer dimension (index $k$) as shown in \figRef{fig:mainFigCgra}.
        {(b)} Then, once the loop indices have been properly determined for the current iteration, the addresses of the matrix elements that are to be accessed in this iteration must be computed.
        This is done by multiplying the loop indices with fixed strides and adding the results together.
        {(c)} Afterward, the computed addresses are used to load the inputs and store the output~(see the memory access section in \figRef{fig:mainFigCgra}.
        A restriction here is that in contrast to the other operations, the corresponding \texttt{Load} and \texttt{Store} operations cannot be executed on all PEs, but only on those PEs that have access to the SPM, which are, as shown in \figRef{fig:mainFigCgra}, only the border PEs.
        {(d)} Only then can the \texttt{Mul} and \texttt{Add} operations, forming the only computational part of the loop nest, \ie, one partial product, be computed before the result is written back by a \texttt{Store} operation.
        Note also that the DFG contains multiple performance-constraining cycles, \eg, \texttt{Sel}$\,\to\,$\texttt{Add}$\,\to\,$\texttt{Cmp}$\,\to\,$\texttt{Sel}, inside the indices computation.
        As a consequence, the \texttt{Sel} operation of the next iteration cannot be started before the \texttt{Cmp} and \texttt{Add} operations of the previous iteration are completed.
        Thus, the cycle length determines a minimal possible \II, called the \emph{recurrence minimum initiation interval}~(RecMII).
        Also, the minimal possible \II may be further limited by a \emph{resource minimum initiation interval}~(ResMII).
        For example, given a CGRA with 9~PEs, the actual minimal possible \II is 3, because with $\II = 2$, each iteration would only allow for $9 \cdot 2 = 18$ nodes to be scheduled.

        \begin{figure*}
            \centering
            \begin{minipage}[c]{0.5\textwidth}%
            \resizebox{1.0\textwidth}{!}{
                \begin{tikzpicture}
                    \definecolor{darkPurple}{RGB}{96, 25, 134}
                    \definecolor{darkBlue}{RGB}{0, 59, 111}
                    \definecolor{darkBrown}{RGB}{102, 51, 0}
                    \definecolor{darkGray}{RGB}{71, 79, 82}
                    \definecolor{darkRed}{RGB}{177, 0, 18}
                    \definecolor{darkGreen}{RGB}{0, 147, 146}
                    \definecolor{darkOrange}{RGB}{111, 59, 0}

                    \pic at (0cm, 0cm) {nodeGrid={1.25cm}{1.05cm}{0.4cm}{
                        {{Sel/s0/white/darkBlue},{},{},{},{Mul/ls0/white/darkOrange},{},{},{Load/load0/white/darkPurple},{},{},{}},
                        {{},{Add/a0/white/darkBlue},{},{},{},{Add/addr0/white/darkOrange},{},{},{},{},{Add/mac0/white/darkRed}},
                        {{Cmp/c0/white/darkBlue},{},{Sel/s1/white/darkBlue},{},{Mul/ls1/white/darkOrange},{},{},{Store/store/white/darkPurple},{},{},{}},
                        {{},{Add/a1/white/darkBlue},{},{},{},{Add/addr1/white/darkOrange},{},{Load/load1/white/darkPurple},{},{},{}},
                        {{},{},{Cmp/c1/white/darkBlue},{},{Mul/ls2/white/darkOrange},{},{},{},{},{},{Mul/mac1/white/darkRed}},
                        {{},{Add/a2/white/darkBlue},{},{},{},{Add/addr2/white/darkOrange}, {}, {Load/load2/white/darkPurple},{},{},{}},
                        {{Cmp/c2/white/darkBlue},{},{Sel/s2/white/darkBlue},{},{Mul/ls3/white/darkOrange},{},{},{},{},{},{}}%
                    }{0.07cm}{
                        {s0/a0/darkBlue//-22.5/115.5/1/{}},
                        {a0/s0/darkBlue//160.5/-67.5/1/{dashed}},
                        {a0/c0/darkBlue//{}},
                        {c0/s0/darkBlue///{dashed}},
                        {s1/a1/darkBlue//-22.5-90/115.5-90/1/{}},
                        {a1/s1/darkBlue//160.5-90/-67.5-90/1/{dashed}},
                        {a1/c1/darkBlue//{}},
                        {c1/s1/darkBlue///{dashed}},
                        {a2/s2/darkBlue//-22.5/115.5/1/{dashed}},
                        {s2/a2/darkBlue//160.5/-67.5/1/{}},
                        {a2/c2/darkBlue//{}},
                        {c2/s2/darkBlue///{dashed}},
                        {c0/a1/darkBlue//{}},
                        {c1/a2/darkBlue//{}},
                        {s0/ls0/darkBlue/i//{}},
                        {s1/ls1/darkBlue/j//{}},
                        {s2/ls2/darkBlue/k//{}},
                        {s2/ls3/darkBlue/k//{}},
                        {ls0/addr0/darkOrange///{}},
                        {ls0/addr1/darkOrange///{}},
                        {ls1/addr0/darkOrange///{}},
                        {ls1/addr2/darkOrange///{}},
                        {ls2/addr1/darkOrange///{}},
                        {ls3/addr2/darkOrange///{}},
                        {addr0/load0/darkOrange///{}},
                        {addr0/store/darkOrange///{}},
                        {addr1/load1/darkOrange///{}},
                        {addr2/load2/darkOrange///{}},
                        {load0/mac0/darkPurple/{C[i, j]}//{}},
                        {load1/mac1/darkPurple/{A[i, k]}//{}},
                        {load2/mac1/darkPurple/{B[k, j]}//{}},
                        {mac0/store/darkRed/{C[i, j]}//{}},
                        {mac1/mac0/darkRed///{}},
                    }{%
                        {4/0/4/8/{dashed}},
                        {7/0/7/8/{dashed}},
                        {10/0/10/8/{dashed}},
                    }{%
                        {0/7/4/8/{Indices\\ Computation}/},
                        {4/7/7/8/{Address\\ Computation}/},
                        {7/7/10/8/{Memory\\ Access}/},
                        {9.5/7/12.5/8/{MAC\\ Operations}/},
                    }};
                \end{tikzpicture}
            }%
            \end{minipage}%
            \begin{minipage}[c]{0.5\textwidth}%
            \resizebox{1.0\textwidth}{!}{
                \begin{tikzpicture}
                    \definecolor{darkPurple}{RGB}{96, 25, 134}
                    \definecolor{darkBlue}{RGB}{0, 59, 111}
                    \definecolor{darkBrown}{RGB}{102, 51, 0}
                    \definecolor{darkGray}{RGB}{71, 79, 82}
                    \definecolor{darkRed}{RGB}{177, 0, 18}
                    \definecolor{darkGreen}{RGB}{0, 147, 146}
                    \definecolor{darkOrange}{RGB}{111, 59, 0}

                    \pic at (0cm, -0.5cm-5*2.00cm) {cgra={4}{4}{1.5cm}{0.5cm}};
                    \pic at (1.5cm, -0.5cm) {nodeGrid={0.5cm}{0.4cm}{0.3cm}{
                        {{},{},{},{},{},{},{},{},{},{},{}},
                        {{},{L/load0/white/darkPurple},{},{},{},{},{},{},{},{S/s0/white/darkBlue},{},{A/a0/white/darkBlue}},
                        {{},{},{},{},{},{},{},{},{},{},{},{}},
                        {{},{S/store/white/darkPurple},{},{},{},{A/addr0/white/darkOrange},{},{},{},{},{C/c0/white/darkBlue},{},{}},
                        {{},{},{},{},{},{},{},{},{},{},{}},
                        {{},{},{},{A/mac0/white/darkRed},{},{},{},{M/ls0/white/darkOrange},{},{S/s1/white/darkBlue},{},{A/a1/white/darkBlue}},
                        {{},{},{},{},{},{},{},{M/ls1/white/darkOrange},{},{},{}},
                        {{},{},{},{M/mac1/white/darkRed},{},{},{},{M/ls2/white/darkOrange},{},{},{C/c1/white/darkBlue},{}},
                        {{},{},{},{},{},{},{},{},{},{},{}},
                        {{},{L/load1/white/darkPurple},{},{},{},{A/addr1/white/darkOrange},{},{},{},{S/s2/white/darkBlue},{},{A/a2/white/darkBlue}},
                        {{},{},{},{},{},{},{},{},{},{},{}},
                        {{},{L/load2/white/darkPurple},{},{},{},{A/addr2/white/darkOrange},{},{M/ls3/white/darkOrange},{},{},{C/c2/white/darkBlue},{}},
                        {{},{},{},{},{},{},{},{},{},{},{}},
                        {{},{},{},{},{},{},{},{},{},{},{},{},{},{},{},{}},
                        {{},{},{},{},{},{},{},{},{},{},{}},
                        {{},{},{},{},{},{},{},{},{},{},{},{},{},{}}%
                    }{0.05cm}{
                        {s0/a0/darkBlue//45/135/1/{}},
                        {a0/s0/darkBlue//225/-45/1/{dashed}},
                        {a0/c0/darkBlue//{}},
                        {c0/s0/darkBlue///{dashed}},
                        {s1/a1/darkBlue//45/135/1/{}},
                        {a1/s1/darkBlue//225/-45/1/{dashed}},
                        {a1/c1/darkBlue//{}},
                        {c1/s1/darkBlue///{dashed}},
                        {s2/a2/darkBlue//45/135/1/{}},
                        {a2/s2/darkBlue//225/-45/1/{dashed}},
                        {a2/c2/darkBlue//{}},
                        {c2/s2/darkBlue///{dashed}},
                        {c0/a1/darkBlue//{}},
                        {c1/a2/darkBlue//{}},
                        {s0/ls0/darkBlue///{}},
                        {s1/ls1/darkBlue///{}},
                        {s2/ls2/darkBlue///{}},
                        {s2/ls3/darkBlue///{}},
                        {ls0/addr0/darkOrange///{}},
                        {ls0/addr1/darkOrange///{}},
                        {ls1/addr0/darkOrange///{}},
                        {ls1/addr2/darkOrange///{}},
                        {ls2/addr1/darkOrange///{}},
                        {ls3/addr2/darkOrange///{}},
                        {addr0/load0/darkOrange///{}},
                        {addr0/store/darkOrange///{}},
                        {addr1/load1/darkOrange///{}},
                        {addr2/load2/darkOrange///{}},
                        {load0/mac0/darkPurple///{}},
                        {load1/mac1/darkPurple///{}},
                        {load2/mac1/darkPurple///{}},
                        {mac0/store/darkRed///{}},
                        {mac1/mac0/darkRed///{}},
                    }{}{}};
                \end{tikzpicture}
            }%
            \end{minipage}
            \caption{%
                A simplified data flow graph~(DFG) of a matrix multiplication~(shown left) is mapped onto a 4$\times 4$ CGRA architecture~(shown right).
            }
            \label{fig:mainFigCgra}
            \vspace{-.7em}
        \end{figure*}
\section{Mapping Tools}
\label{sec:tools}
        Many different CGRA architectures have been proposed over the years, and some related toolchains are also publicly available.
        This study selects four representative toolchains for analysis, briefly introduced in the following.
        \subsubsection{CGRA-Flow} \label{sec:cgra:tools:cgraflow}
            CGRA-Flow~\cite{4_OpenCGRA}, also known as Open\-CGRA, is a toolchain for the compilation, exploration, synthesis, and development of CGRA architectures~\cite{4_OpenCGRA}.
            It is open-source and available on GitHub\footnote{\url{https://github.com/tancheng/CGRA-Flow}}.
            CGRA-Flow has a GUI for visualizing input, output, and intermediate results.
            As input, users describe or select a loop program written in C/C++.
            CGRA-Flow supports up to two innermost loop nests with control flow in the loop body or up to three innermost loop nests without control flow in the loop body onto the user-specified CGRA.
            Within the GUI, users can configure a CGRA architecture instance by selecting the number of PEs, the number of operations mapped to one PE, the size of the memory buffer, the operation types that the PE can execute, the connections to neighboring PEs, and the disablement of entire PEs.
            Compiling the user-given loop and generating the corresponding DFG is performed using LLVM's~\cite{LLVM} intermediate representation to extract the operations and the dependency between operations.
            The user can select between two mapping algorithms, called \emph{exhaustive} and \emph{heuristic}.
            While the exhaustive algorithm checks all possible mappings for one given initiation interval \II, the heuristic approach starts with a minimal initiation interval and iteratively increments it until a mapping with the lowest cost (based on a heuristic function) for the current initiation interval \II has been found.
            Both the generated DFG and the resulting mapping are visualized in the GUI.
            After mapping, the user can generate Verilog via PyMTL to run various tests on the architecture~\cite{4_OpenCGRA} and to estimate the area and power of PEs and on-chip memory.
        \subsubsection{Morpher}\label{sec:cgra:tools:morpher}
            Morpher is an integrated compilation and simulation toolchain~\cite{6_MorpherWOSET} available on GitHub\footnote{\url{https://github.com/ecolab-nus/morpher}}.
            As input, the user provides a description of the target CGRA architecture and a loop program written in C/C++ that should be mapped onto the target architecture.
            The DFG generator begins by extracting the innermost loop of the program and generating the corresponding DFG using LLVM~\cite{LLVM}.
            It offers three schemes for handling loop control flow in a CGRA, influencing DFG generation: \emph{partial predication}, \emph{full predication}, and \emph{dual-issue}~\cite{6_MorpherWOSET}, described in detail in~\cite{8_Branch_aware_loop_mapping}.
            Partial predication maps the if-part and else-part operations to different PEs, adding a \emph{select} node if both parts update the same variable.
            Full predication schedules both parts using the same variable to the same PE, with one operation executed per cycle, avoiding the need for a select node.
            Dual-issue merges both operations into one DFG node, scheduling them simultaneously but only executing one at run-time.
            In this work, we only consider partial predication because it was the most reliably supported mapping.
            The resulting DFG and data layout for input/output variables on the SPM are then used by the CGRA mapper to find a valid mapping using three algorithms: PathFinder~\cite{McMurchieE95}, Simulated Annealing~\cite{32_SA}, and LISA~\cite{LiMitra2022HPCA}.
            The mapping can then be verified by simulating the execution using a simulator that models a CGRA with FUs, registers, multiplexers, and memory banks~\cite{6_MorpherWOSET}, supporting variations of HyCUBE.
            Here, the test data is automatically created by Morpher’s data generator~\cite{3_MorpherCODAI}.
            After simulation, memory values are compared to the test data.
        \subsubsection{Pillars}\label{sec:cgra:tools:pillars}
            Pillars is an open-source CGRA design toolchain based on Scala and Chisel~\cite{guo-pillars-woset2020}.
            The toolchain is publicly available on GitHub\footnote{\url{https://github.com/pku-dasys/pillars}}.
            It has been designed as a tool for conducting design space explorations and further hardware optimizations of CGRAs.
            The user must provide a Scala-based architecture description of the CGRA, which is then systematically converted by Chisel into a synthesizable Verilog description of the CGRA.
            The design can be synthesized for FPGAs to determine the performance, area and power consumption of the CGRA.
            In contrast to the other toolchains, the Pillars toolchain does not support any automated DFG generation from source code.
            Thus, the user must already provide the application as a DFG.
            Pillars offers two mapping algorithms, an ILP mapper and a Heuristic Search mapper.
            The ILP mapper is slow but succeeds more frequently than the Heuristic Search mapper.
            Therefore, the ILP mapper is used for all loop kernels in this work.
            The resulting mapping can then be either simulated on the RTL by Verilator, or used to determine the performance, area and power estimate for the execution on the actual hardware.
        \subsubsection{CGRA-ME}\label{sec:cgra:tools:cgrame}
            CGRA-ME is an open-source toolchain for modeling and exploration of CGRAs~\cite{RaghebWWBRYA24}.
            It is currently available in its second version~(first release~\cite{17_CGRA-ME}) and supports end-to-end CGRA compilation and simulation with RTL code generation.
            The toolchain is open-source and can be downloaded from the project's website\footnote{\url{https://cgra-me.ece.utoronto.ca/download/}}.
            It uses LLVM to extract a DFG from a given C/C++ source code.
            However, it does not support any nested loops as inputs, but although it supports partial predication, no support for conditional code was available.
            CGRA-ME maps the extracted DFG onto a target CGRA, which is specified by a provided architecture description.
            It offers to choose between three different mapping approaches.
            First, CGRA-ME supports an ILP-based mapping that finds an optimal mapping but is very slow.
            Additionally, a heuristic approach reduces the search space of the ILP.
            Finally, CGRA-ME also includes a clustered mapper that incorporates a simulated-annealing approach~\cite{32_SA} utilizing both QuickRoute~\cite{LiE04} and PathFinder~\cite{McMurchieE95}.
            After the mapping, CGRA-ME produces a Verilog description of the given architecture and a bitstream containing the configuration of the CGRA.
            However, simulation without additional external toolchains is not available.
\section{Comparative Evaluation}
\label{sec:evaluation}
    \begin{table}
    \caption{%
        Mapping results of benchmarks on CGRAs.
    }
    \label{table:tcpa_cgra_mapping}
\hspace{-3mm}
    \resizebox{0.5\textwidth}{!}{
        \definecolor{darkGray}{RGB}{71, 79, 82}   
        \definecolor{darkRed}{RGB}{177, 0, 18}
        \definecolor{darkOrange}{RGB}{207, 112, 4}
        \rowcolors{1}{darkGray!20}{darkGray!5}
        \begin{tabular}{|l|c|c|c|c|c|c|}
          \hline 
 \rowcolor{darkGray} \color{white}{\textbf{Toolchain}} & \color{white}{\textbf{Optimization}} & \color{white}{\textbf{Architecture}} & \color{white}{\textbf{\#op.}} & \color{white}{\textbf{\II}} & \color{white}{\textbf{\#unused PE}} & \color{white}{\textbf{max(\#op. per PE)}} \\\hline\hline
 \rowcolor{darkGray} \multicolumn{7}{|c|}{\color{white}{\textbf{GEMM}}} \\\hline
 {CGRA-Flow}                                          & -                                    & cls. CGRA                            & 23                            & 10                         & 6                                   & 4 \\ 
 {CGRA-Flow}                                          & flat                                 & cls. CGRA                            & 28                            & 6                          & 5                                   & 4 \\ 
 {CGRA-Flow}                                          & flat+unroll                          & cls. CGRA                            & 52                            & 6                          & 0                                   & 6 \\
 {Morpher}                                            & {flat}                               & cls. CGRA                            & 47                            & 9                          & 4                                   & 9 \\ 
 {Morpher}                                            & {flat}                               & HyCUBE                               & 47                            & 9                          & 5                                   & 9 \\ 
 {Morpher}                                            & {flat+unroll}                        & cls. CGRA                            & 80                            & 8                          & 2                                   & 8 \\ 
 {Morpher}                                            & {flat+unroll}                        & HyCUBE                               & 80                            & 8                          & 1                                   & 8 \\
 \rowcolor{darkOrange!50}{CGRA-ME}                    & {-}                                  & HyCUBE                               & 23                            & 1                          & 5                                   & 1 \\
 \rowcolor{darkOrange!50}{Pillars}                    & {-}                                  & ADRES                                & 23                            & 1                          & 5                                   & 1 \\\hline\hline
 \rowcolor{darkGray} \multicolumn{7}{|c|}{\color{white}{\textbf{ATAX}}} \\\hline
 {CGRA-Flow}                                          & -                                    & cls. CGRA                            & 30                            & 13                         & 3                                   & 6 \\ 
 {CGRA-Flow}                                          & flat                                 & cls. CGRA                            & 36                            & 10                         & 0                                   & 4 \\
 {CGRA-Flow}                                          & flat+unroll                          & cls. CGRA                            & 87                            & 25                         & 0                                   & 8 \\ 
 {Morpher}                                            & {flat}                               & cls. CGRA                            & 55                            & 14                         & 3                                   & 10 \\
 {Morpher}                                            & {flat}                               & HyCUBE                               & 55                            & 10                         & 1                                   & 8 \\
 \rowcolor{darkRed!50}{Morpher}                       & {flat+unroll}                        & cls. CGRA                            & 118                           & -                          & -                                   & - \\ 
 {Morpher}                                            & {flat+unroll}                        & HyCUBE                               & 118                           & 14                         & 0                                   & 14 \\
 \rowcolor{darkOrange!50}{CGRA-ME}                    & {-}                                  & HyCUBE                               & 21                            & 2                          & 11                                  & 2 \\
 \rowcolor{darkRed!50}{Pillars}                       & {-}                                  & ADRES                                & 21                            & -                          & -                                   & - \\\hline\hline
 \rowcolor{darkGray} \multicolumn{7}{|c|}{\color{white}{\textbf{GESUMMV}}} \\\hline
 {CGRA-Flow}                                          & -                                    & cls. CGRA                            & 22                            & 8                          & 6                                   & 4\\ 
 {CGRA-Flow}                                          & flat                                 & cls. CGRA                            & 25                            & 5                          & 3                                   & 4 \\ 
 {CGRA-Flow}                                          & flat+unroll                          & cls. CGRA                            & 58                            & 7                          & 0                                   & 6 \\
 {Morpher}                                            & {flat}                               & cls. CGRA                            & 41                            & 6                          & 4                                   & 6 \\ 
 {Morpher}                                            & {flat}                               & HyCUBE                               & 41                            & 6                          & 3                                   & 6 \\
 \rowcolor{darkRed!50}{Morpher}                       & {flat+unroll}                        & cls. CGRA                            & 86                            & -                          & -                                   & - \\ 
 {Morpher}                                            & {flat+unroll}                        & HyCUBE                               & 86                            & 9                          & 2                                   & 6 \\ 
 \rowcolor{darkOrange!50}{CGRA-ME}                    & {-}                                  & HyCUBE                               & 28                            & 7                          & 10                                  & 3 \\
 \rowcolor{darkRed!50}{Pillars}                       & {-}                                  & ADRES                                & 28                            & -                          & -                                   & - \\\hline\hline
 \rowcolor{darkGray} \multicolumn{7}{|c|}{\color{white}{\textbf{MVT}}} \\\hline
 {CGRA-Flow}                                          & -                                    & cls. CGRA                            & 29                            & 8                          & 3                                   & 5 \\ 
 {CGRA-Flow}                                          & flat                                 & cls. CGRA                            & 31                            & 5                          & 2                                   & 4 \\
 {CGRA-Flow}                                          & flat+unroll                          & cls. CGRA                            & 73                            & 7                          & 0                                   & 6 \\
 {Morpher}                                            & {flat}                               & cls. CGRA                            & 49                            & 7                          & 3                                   & 7 \\ 
 {Morpher}                                            & {flat}                               & HyCUBE                               & 49                            & 7                          & 3                                   & 7\\ 
 {Morpher}                                            & {flat+unroll}                        & cls. CGRA                            & 106                           & 9                          & 0                                   & 9 \\ 
 {Morpher}                                            & {flat+unroll}                        & HyCUBE                               & 106                           & 8                          & 0                                   & 8 \\ 
 \rowcolor{darkOrange!50}{CGRA-ME}                    & {-}                                  & HyCUBE                               & 21                            & 2                          & 11                                  & 2 \\
 \rowcolor{darkRed!50}{Pillars}                       & {-}                                  & ADRES                                & 21                            & -                          & -                                   & - \\\hline\hline
 \rowcolor{darkGray} \multicolumn{7}{|c|}{\color{white}{\textbf{TRISOLV}}} \\\hline
 {CGRA-Flow}                                          & -                                    & cls. CGRA                            & 27                            & 10                         & 3                                   & 4\\ 
 \rowcolor{darkRed!50}{CGRA-Flow}                     & flat                                 & cls. CGRA                            & 44                            & -                          & -                                   & -\\ 
 \rowcolor{darkRed!50}{CGRA-Flow}                     & flat+unroll                          & cls. CGRA                            & 138                           & -                          & -                                   & -\\ 
 {Morpher}                                            & {flat}                               & cls. CGRA                            & 57                            & 8                          & 3                                   & 7 \\ 
 {Morpher}                                            & {flat}                               & HyCUBE                               & 57                            & 7                          & 3                                   & 4 \\
 \rowcolor{darkRed!50}{Morpher}                       & {flat+unroll}                        & cls. CGRA                            & 180                           & -                          & -                                   & -\\
 \rowcolor{darkRed!50}{Morpher}                       & {flat+unroll}                        & HyCUBE                               & 180                           & -                          & -                                   & -\\ 
 \rowcolor{darkOrange!50}{CGRA-ME}                    & {-}                                  & HyCUBE                               & 21                            & 2                          & 11                                  & 2 \\
 \rowcolor{darkRed!50}{Pillars}                       & {-}                                  & ADRES                                & 21                            & -                          & -                                   & - \\ \hline
        \end{tabular}
    }
    \vspace{-1.5em}
\end{table}

    In this section, we evaluate the mapping quality each of the discussed mapping tools achieved.
    For evaluation, we picked five common loop benchmarks (GEMM, ATAX, GESUMMV, MVT, and TRISOLV) from the Polybench suite~\cite{polybench}.
    Each benchmark is a multidimensional loop nest representing a typical workload in the linear algebra domain.
    We observed that several CGRA toolchains do not accept as input a multidimensional loop nest directly, but require flattening, \ie, the multidimensional loop nest is reduced into a single loop by unfolding the iterations of the outer loops.
    Furthermore, no considered CGRA toolchain unrolls a given loop automatically; thus, this transformation was done manually.
    Then, we mapped each benchmark kernel using the four selected CGRA toolchains: CGRA-Flow~\cite{4_OpenCGRA}, Morpher~\cite{6_MorpherWOSET}, CGRA-ME~\cite{RaghebWWBRYA24}, and Pillars~\cite{guo-pillars-woset2020}.
    \tableRef{table:tcpa_cgra_mapping} summarizes the mapping results.
    Whenever no mapping could be found, the entry is colored red, while an orange row indicates a successful mapping of only the innermost loop.
    CGRA-Flow is the only CGRA tool that directly supports multidimensional loops.
    However, we can observe that flattening should still be applied in all but one benchmark as it favorably reduces the achievable initiation interval.
    Only in the case of TRISOLV, the flattened benchmark could not be mapped by CGRA-Flow.
    In this evaluation, we used two different target architectures for Morpher.
    A classical CGRA without multi-hop connections and HyCube, for which Morpher finds consistently better mappings.
    We targeted an ADRES-like~\cite{MeiVVML03} architecture for Pillars, and since Pillars does not come with its own DFG generator, we utilized the DFG from CGRA-ME.
    However, it could only find a valid mapping for the GEMM kernel, failing for all others.
    Without any direct support for multidimensional loops, the loop must be flattened into a single loop.
    This requires explicitly inserting conditional statements inside the loop body that update the outer loop indices.
    Moreover, CGRA-ME currently does not support any predication; hence, it only maps the innermost loop.
    But this simplification allows this tool to achieve the lowest initiation interval \II among the CGRA toolchains.
    However, as discussed in \secRef{sec:cgra:mapping}, the generation of the loop indices should introduce a RecMII of 3.
    This does not apply to CGRA-ME because it only maps the innermost loop and omits any loop-bound checks.
    Besides the achieved initiation interval, the table also shows the number of PEs that were not utilized and the maximum number of operations mapped to a PE.
    It was rarely possible to employ all the 16 available PEs.
    Moreover, the number of operations per PE indicates that even the used PEs are underutilized.
    For example, with $\II=10$ and a maximum of 4 operations per PE, the most active PE will only execute 4 operations within 10 cycles.

\section{Summary and Outlook}
\label{sec:summary}
    In this paper, we evaluated and examined several coarse-grained reconfigurable arrays (CGRAs) and related toolchains that directly map operations from a data flow graph (DFG) to processing elements (PEs).
    We mapped four common loop benchmarks using four publicly available CGRA tools and observed that in many resulting mappings, multiple PEs stay inactive, \ie, no operations are mapped to them, while the maximal number of operations mapped on a PE further indicate that those PEs are not fully utilized.
    Thus, these mapping approaches still have headroom to utilize the available array of PEs fully.
    This is likely to be the case because of the limited routing capability of the PEs, which makes it impossible for every PE to execute an operation in every cycle.
    We can also observe that the HyCube architecture, which increases the routing capability by introducing multi-hop connections, consistently outperforms the classic CGRAs with no multi-hop connections.
    Finally, we observed that Morpher beats CGRA-Flow in all but one benchmark.
    Although CGRA-ME shows the lowest \II in all benchmarks, it is important to note that this includes only the innermost loop.
    Pillars, however, fails in almost all benchmarks.
    In further work~\cite{walter2025mappingexecutionnestedloops}, we extend this research to a more comprehensive evaluation of processor arrays, considering both qualitative and quantitative measures.
    This includes not only CGRAs but also another contender for loop acceleration---so-called Tightly-Coupled Processor Arrays~(TCPAs)~\cite{KisslerTeich2006IEEEFPT, HannigReiche2014ACMTECS, TeichWitterauf2022Book}, which differ significantly in their mapping methodology~\cite{TeichT93, WitteraufWHT21}.

\vspace{5pt}
{\small%
\noindent%
\textbf{Acknowledgment:} This work was partly funded by the Deutsche Forschungsgemeinschaft (DFG, German Research Foundation) -- Project number 146371743 -- TRR 89: Invasive Computing.%
}

\renewcommand*{\bibfont}{\scalefont{0.925}}
\footnotesize
\printbibliography

\end{document}